\documentclass[12pt]{article}
\usepackage[centertags]{amsmath}
\usepackage{algorithm}
\usepackage{algpseudocode}
\usepackage{amsthm}
\usepackage{float}
\usepackage{newlfont}
\usepackage{amsmath}
\usepackage{amsfonts,amsmath,latexsym}
\usepackage{mathrsfs}
\usepackage{amssymb}
\usepackage{amsbsy}
\usepackage{indentfirst}
\usepackage{mathrsfs}
\usepackage{subfig}
\usepackage{graphicx}
\usepackage{xcolor}
\usepackage{amssymb, amsmath}
\usepackage{bm}
\usepackage{wasysym}
\usepackage[flushleft]{threeparttable}
\usepackage{natbib}

\usepackage{hyperref}
\hypersetup{
    colorlinks = true,
    allcolors = {blue}
}

\usepackage{diagbox}
\usepackage{rotating}
\usepackage{booktabs}
\usepackage{bigstrut, multirow}
\allowdisplaybreaks[4]

\titlepage
\newlength{\defbaselineskip}
\newcommand{\be}{\begin{eqnarray}}
	\newcommand{\ee}{\end{eqnarray}}
\newcommand{\bestar}{\begin{eqnarray*}}
	\newcommand{\eestar}{\end{eqnarray*}}
\newcommand{\ignore}[1]{}
{} \theoremstyle{plain}
\newtheorem{thm}{THEOREM}[section]

\newtheorem{lem}{LEMMA}[section]

\theoremstyle{definition}

\newtheorem{assum}{ASSUMPTION}[section]

\theoremstyle{remark}

\numberwithin{equation}{section}

\def\s{\sigma}
\def\S{\Sigma}
\def\W{\Omega}

\def\ct{\theta}
\def\CT{\Theta}

\def\>{\geq}
\def\<{\leq}

\def\mR{\mathbb R}

\usepackage{enumitem}
\newlist{steps}{enumerate}{1}
\setlist[steps, 1]{label = Step \arabic*:}
\def\step{%
   \@ifnextchar[ \@step{\@noitemargtrue\@step[\@itemlabel]}}
\def\@step[#1]{\item[#1]\mbox{}\\\hspace*{\dimexpr-\labelwidth-\labelsep}}
\makeatother
\begin{document}
	\baselineskip=0.85 true cm   
	
	\title{
		Estimation for conditional moment models based on martingale difference divergence
	}
	
	\date{}
	\author{Kunyang Song$^1$, Feiyu Jiang$^2$\footnote{Address correspondence to Department of Statistics and Data Science, School of Management, Fudan University, Shanghai, China.  E-mail: jiangfy@fudan.edu.cn
		} , and Ke Zhu$^1$\\\\
		\emph{University of Hong Kong$^1$ and Fudan University$^2$}
		\date{}
	}

	\maketitle

	\begin{abstract}
	We provide a new estimation method for conditional moment models via the martingale difference divergence (MDD).
    Our MDD-based estimation method is formed in the framework of a continuum of unconditional moment restrictions. Unlike the existing
    estimation methods in this framework, the MDD-based estimation method adopts a non-integrable weighting function, which could grab more information from unconditional moment restrictions than the integrable weighting function to enhance the estimation efficiency. Due to the nature of shift-invariance in MDD, our MDD-based estimation method can not identify the intercept parameters. To overcome this identification issue,
    we further provide a two-step estimation procedure for the model with intercept parameters.
        Under regularity conditions, we establish the asymptotics of the proposed estimators, which are not only easy-to-implement with analytic asymptotic variances, but also applicable to time series data with an unspecified form of conditional heteroskedasticity.
    Finally, we illustrate the usefulness of the proposed estimators by simulations and two real examples.
	\end{abstract}


\noindent{\it MOS subject classification: 62M10, 62M20}
	
	\newpage



	\section{Introduction}\label{sec_1}

Conditional moment models play a pivotal role in various applications within economics and statistics.
Typically, these models assume an unknown parameter $\theta_0$, which is subject to the following conditional moment restrictions \citep{C:1987}:
\begin{equation}\label{model}
E(h(Z_t,\ct_0)|X_t)=0 \,\,\text{almost surely (a.s.),}
\end{equation}
where $\ct_0 \in \CT \subset \mR^d$, $Z_t$ is a $k$-dimensional random vector that may contain both endogenous and exogenous variables, $X_t$ is a $q$-dimensional random vector that contains the conditioning variables, and $h$ is a given smooth function mapping $\mR^k\times\CT$ into $\mR^l$.
To estimate $\ct_0$ in (\ref{model}), the conventional approach is to employ unconditional moment restrictions, such as the
generalized method of moments (GMM) (\citealp{Hansen:1982}). To be specific, by transforming conditional moment restrictions in (\ref{model}) to unconditional ones,  we can construct estimating moment functions $E(f(X_t)h(Z_t,\ct_0))=0$ for a $r\times l$
matrix of instrumental variables (IVs) $f(X_t)$. To attain the semiparametric efficiency bound,
\citet{Newey:1990,Newey:1993} choose the optimal IVs by using the nearest neighbors or series
expansions method. See the similar non-parametric implementation in \citet{DIN:2003} and \citet{HK:2011}, and its further development on variational method of moments in \citet{BK:2003} for univariate $X_t$. Notwithstanding the efficiency of the aforementioned estimators, they have three drawbacks that possibly hinder their practical application.
First, their asymptotic theory is only established for the independent and identically distributed (i.i.d.) sample $\{(X_t, Z_t)\}_{t=1}^{n}$.
Second, their implementation requires the use of tuning parameters or a well-behaved preliminary estimator, or sometimes both.
Third, despite the use of optimal IVs, they may fail to identify $\theta_0$ when $h(Z_t,\ct_0)$ is a nonlinear function.

To address the above drawbacks, \citet{DL:2004} propose an alternative method for estimating $\theta_0$ by considering a continuum of unconditional moment restrictions. These restrictions
essentially correspond to an infinite number of IVs that can span a space of functions of $X_t$ (\citealp{Bierens:1982,Bierens:1990}).
Specifically,  \citet{DL:2004} consider the restrictions of
\begin{equation}\label{dl_idea}
E(h(Z_t,\ct_0)\mbox{I}(X_t\leq s))=0 \,\,\text{for }s\in\mathbb{R}^{q},
\end{equation}
where $\mbox{I}(\cdot)$ is the indicator function, and $\mbox{I}(X_t\leq s)$ is an IV for each $s$.
To account for the effect of all IVs in (\ref{dl_idea}), they estimate $\theta_0$ by minimizing the sample counterpart of
\begin{equation}
\mbox{DL}(\theta_0)\triangleq\int_{\mathbb{R}^{q}} \|E(h(Z_t,\ct_0)\mbox{I}(X_t\leq s))\|^2 w(s)ds,
\end{equation}
where $\|\cdot\|$ is the Frobenius norm, and the weighting function $w(s)$ is the density function of $X_t$.
See also \citet{Escanciano:2006} for similar ideas in the i.i.d. setting.
It is worth noting that the idea of \citet{DL:2004} has a linkage to that of
\citet{CF:2000}, which uses the exponential function $\exp(sX_t)$ as the IV for
univariate $X_t$ and $s$ in a neighborhood of $0$.
When $\{(X_t, Z_t)\}_{t=1}^{n}$ is an i.i.d. sample, the parameter estimation
using the exponential function to form an infinite number of IVs
is further investigated by \citet{LP:2013} for model (\ref{model}), \citet{Escanciano:2018} for linear models, and
\citet{AS:2022} for partially linear models. When $\{(X_t, Z_t)\}_{t=1}^{n}$ is temporally dependent, an extension study of \citet{Escanciano:2018} can be found in \citet{CEG:2022} for linear models. Regardless whether indicator or exponential function is adopted to generate IVs, these estimation methods relying on a continuum of unconditional moment restrictions necessitate the condition that $w(s)$ is  integrable.
 It is important to note that due to the integrability of $w(s)$, its value likely decays rapidly to zero as $|s|\to\infty$, especially when the dimension of $s$ is large (see \cite{Wang:2024}).  It turns out that the influence of IVs associated with larger values of $s$ is substantially diminished through the use of $w(s)$, bringing an adverse impact on the efficiency of model estimation.

Within the framework of a continuum of unconditional moment restrictions, our paper is motivated by the martingale difference divergence (MDD) introduced by \citet{SZ:2014}. We propose to estimate $\theta_0$ using a non-integrable weighting function  as:
\begin{equation}\label{weighting_function}
w_*(s)=\frac{1}{c_{q}\|s\|^{1+q}} \,\,\text{for }s\in\mathbb{R}^{q},
\end{equation}
where $c_q=\pi^{(1+q)/2}/\Gamma((1+q)/2)$ with $\Gamma(\cdot)$ being the gamma function. The proposed MDD-based estimator relies on the fact that under (\ref{model}),
\begin{equation}\label{eqn1_5}
E\big[h(Z_t,\ct_0)-E(h(Z_t,\ct_0))|X_t\big]=0 \,\,\,\,\text{(a.s.)}.
\end{equation}
From (\ref{eqn1_5}), we consider a continuum of unconditional moment restrictions
\begin{equation}\label{eqn1_6}
\begin{split}
E\big\{[h(Z_t,\ct_0)-E(h(Z_t,\ct_0))]\exp(\mathrm{i} \langle s, X_t\rangle)\big\}
=0 \,\,\text{for }s\in\mathbb{R}^{q},
\end{split}
\end{equation}
where $\mathrm{i}$ is the imaginary unit, $\langle a, b\rangle$ is the inner product for $a, b\in \mathbb{R}^{q}$, and
$\exp(\mathrm{i} \langle s, X_t\rangle)$ is an IV for each $s$.
In view of (\ref{eqn1_6}), the MDD-based estimator aims to minimize the sample counterpart of
\begin{equation}\label{eqn1_7}
\mbox{MDD}(\theta_0)\triangleq\int_{\mathbb{R}^{q}} \Big|E\big\{[h(Z_t,\ct_0)-E(h(Z_t,\ct_0))]\exp(\mathrm{i} \langle s, X_t\rangle)\big\}\Big|_{l}^2 w_*(s)ds,
\end{equation}
where $w_*(s)$ is defined in (\ref{weighting_function}). Here, we denote $|x|_{l}=\sqrt{x^{\star}x}$
for a $l$-dimensional complex vector $x$, with $x^{\star}$ being the conjugate transpose of $x$.
Notably, our MDD-based estimator essentially relies on the conditional moment restrictions in (\ref{eqn1_5}) rather than those in (\ref{model}). Hence, unlike the aforementioned estimators,
it has a different identification condition, which prevents identifying intercept parameters (if exist) for most of models. To deal with this identification issue on intercept parameters, we further propose a two-step estimation procedure for the model with intercept parameters. To be specific, we first estimate all of non-intercept parameters by the MDD-based estimation method at step one. Based on the MDD-based estimator from the step one, we then estimate all of intercept parameters using a straightforward moment estimation approach at step two.

Under regularity conditions that allow for time series sample $\{(X_t, Z_t)\}_{t=1}^{n}$, we provide the consistency and asymptotic normality of the proposed estimator. More importantly, we manage to derive its   asymptotic variance in analytical form, facilitating the  statistical inference procedure.
To demonstrate the effectiveness of our proposed estimators, we conduct extensive simulation studies. The results illustrate a significant advantage in efficiency compared to the estimator proposed in \citet{DL:2004}. This highlights the importance of the non-integrable weighting function $w_*(s)$ to construct parameter estimators for conditional moment time series models. Finally, we demonstrate the superiority of our proposed estimators over the estimator in
\citet{DL:2004} by two real examples.

For univariate linear model with i.i.d. data sample, \citet{Tsyawo:2023} also applies $w_*(s)$ to construct an integrated conditional moment (ICM) estimator by
minimizing the sample counterpart of
\begin{equation}\label{eqn1_8}
\mbox{ICM}(\theta_0)\triangleq\int_{\mathbb{R}^{q}} \Big|E\big\{h(Z_t,\ct_0)\exp(\mathrm{i} \langle s, X_t\rangle)\big\}\Big|_{l}^2 w_*(s)ds,
\end{equation}
rooting in a continuum of unconditional moment restrictions $E\big\{h(Z_t,\ct_0)\exp(\mathrm{i} \langle s, X_t\rangle)\big\}
=0$ for $s\in\mathbb{R}^{q}$. In view of (\ref{eqn1_7})--(\ref{eqn1_8}), the MDD-based estimator is different from the ICM estimator by
subtracting $E(h(Z_t,\ct_0))$ from $h(Z_t,\ct_0)$ to create unconditional moment restrictions.
Due to the additional substraction, the sample counterpart of $\mbox{MDD}(\theta_0)$ in (\ref{eqn1_7}) has an attractive
integral form. This is key to
deriving our asymptotics for time series data sample, as it enables the use of proof technique for the weak convergence in the functional space. However, such a crucial integral form does not exist for the sample counterpart of $\mbox{ICM}(\theta_0)$ in (\ref{eqn1_8}),
and \citet{Tsyawo:2023} proves the asymptotics of the ICM estimator by using a different proof technique that possibly only works for univariate linear model with i.i.d. data sample.
Although the substraction makes the MDD-based estimator applicable for general time series models, it
leads to an aforementioned identification issue on intercept parameters as a trade off. Owing to our well-shaped two-step estimation procedure, this identification issue is practically irrelevant in most of cases. Hence, our MDD-based estimation method could have a much larger application scope than the ICM estimation method in \citet{Tsyawo:2023} to deal with the multiple linear/nonlinear time series models, which are allowed to have an unspecified form of conditional heteroskedasticity.

The remaining paper proceeds as follows. Section \ref{sec_2}
introduces the MDD-based estimator and establishes its asymptotics. Section \ref{sec_3} provides the two-step estimation procedure and its related asymptotics. Simulation results are reported in Section \ref{sec_4}.
Two real examples are given in Section \ref{sec_5}. Concluding remarks are offered in Section \ref{sec_6}. Technical proofs are presented in the Appendix.

\section{The MDD-based Estimator and its Asymptotics}\label{sec_2}

\subsection{The MDD-based Estimator}

For any two vectors $V\in\mathbb{R}^{l}$ and
$U\in\mathbb{R}^{q}$, the MDD of $V$ given $U$ (denoted by $\mbox{MDD}(V|U)$) in \citet{SZ:2014} is defined as follows:
\begin{flalign*}
\operatorname{MDD}(V|U)^{2}&\triangleq\int_{\mathbb{R}^{q}} \Big|g_{V, U}(s)-g_{V} g_{U}(s)\Big|_{l}^{2}w_*(s) ds \\
&=\int_{\mathbb{R}^{q}} \Big|E\big\{[V-E(V)]\exp(\mathrm{i} \langle s, U\rangle)\big\}\Big|_{l}^{2}w_*(s) ds,
\end{flalign*}
where $w_*(s)$ is defined in (\ref{weighting_function}), $g_{V, U}(s)=E\big(V \exp(\textrm{i} \langle s, U\rangle)\big)$, $g_{V}={E}(V)$, and $g_{U}(s)=E\big(\exp(\textrm{i}\langle s, U\rangle)\big)$.  \citet{SZ:2014} show that
$\operatorname{MDD}(V|U)^{2}$ has an equivalent expression as
 \begin{flalign}\label{MDD_form}
\operatorname{MDD}(V|U)^{2}=-E\Big\{\big[V-E(V))^{\top}(V'-E(V')\big]\|U-U'\|\Big\},
\end{flalign}
where $(V', U')$ is an i.i.d. copy of $(V, U)$, and $A^{\top}$ is the transpose of a matrix $A$.
Hence, $\mbox{MDD}(\theta_0)$ in (\ref{eqn1_7}) can be re-written as
 \begin{flalign*}
 \begin{split}
\mbox{MDD}(\theta_0)&=\operatorname{MDD}(h(Z_t,\theta_0)|X_t)^{2}\\
&=-E\Big\{\big[\big(h(Z_t,\theta_0)-E(h(Z_t,\theta_0))\big)^{\top}\big(h(Z_{t'},\theta_0)-E(h(Z_{t'},\theta_0))\big)\big]\|X_t-X_{t'}\|\Big\},
\end{split}
\end{flalign*}
where $(Z_{t'}, X_{t'})$ is an i.i.d. copy of $(Z_t, X_t)$. According to the preceding formula, $\mbox{MDD}(\theta_0)$ has the following sample counterpart:
 \begin{flalign}\label{mdd_samle_form}
 \begin{split}
\mbox{MDD}_n(\theta)&=-\frac{1}{n^2}
\sum_{t=1}^{n}\sum_{t'=1}^{n}
\Big\{\big[\big(h(Z_t,\theta)-\bar{h}(\theta)\big)^{\top}\big(h(Z_{t'},\theta)-\bar{h}(\theta)\big)\big]\|X_t-X_{t'}\|\Big\},
\end{split}
\end{flalign}
where $\bar{h}(\theta)=n^{-1}\sum_{t=1}^{n}h(Z_t,\theta)$. Then, our MDD-based estimator of $\theta_0$ (denoted by $\widehat{\theta}_n$) is defined as the minimizer of $\mbox{MDD}_n(\theta)$ for $\theta\in\Theta$, that is,
\begin{equation*}
\widehat{\theta}_n=\mathop{\arg\min}\limits_{\ct \in \CT}\operatorname{MDD}_{n}(\theta),
\end{equation*}
where $\Theta\subset \mathbb{R}^{d}$ is a parameter space.


\subsection{Asymptotic Properties}

Let $X_t$ and $Z_t$ in (\ref{model}) be two subvectors of $Y_t$. We make the following assumptions to
derive the asymptotics of $\widehat{\theta}_n$.

\begin{assum}
\label{hatii}
$E\big[h(Z_t,\ct)-E(h(Z_t,\ct))|X_t\big]=0$ (a.s.) if and only if $\ct=\ct_0$.
\end{assum}

\begin{assum}
\label{stationarity}
$Y_t$ is strictly stationary and ergodic.
\end{assum}

\begin{assum}
\label{c2}
(i) $h(z,\theta)$ is twice continuously differentiable for each $(z,\theta) \in \mathbb{R}^{k}\times \Theta$;
(ii) there exists a function $g(z)$ such that $\|\partial h(z,\theta)/\partial\theta\|\leq g(z)$ and $E\|g(Z_t)\|^4<\infty$;
(iii) $E\|X_t\|^4 <  \infty$ and $E\prod_{i=1}^q \|X_{it}\|^{2u_0} < \infty$ for some $u_0>1$, where $X_{it}$ is the $i$th entry of
$X_t$.
\end{assum}

\begin{assum}
\label{interpoint}
$\CT$ is compact and $\ct_0$ is an interior point of $\CT$.
\end{assum}

\begin{assum}
\label{whitenoise}
$h(Z_t,\ct_0)$ is a martingale difference sequence with respect to the filtration $\mathcal{F}_t=\s(Y_s,s\leq t)$.
\end{assum}


Our assumptions above are similar to those in \citet{DL:2004}, except that we need a different identification condition for $\theta_0$ in Assumption \ref{hatii}. For the estimator in \citet{DL:2004} or other existing conditional moment estimators, the identification condition for $\theta_0$ is
\begin{flalign}\label{iden_cond_normal}
E(h(Z_t,\ct)|X_t)=0 \mbox{ (a.s.) if and only if }\ct=\ct_0.
\end{flalign}
To illustrate the distinction between the identification conditions in Assumption \ref{hatii} and (\ref{iden_cond_normal}), we
consider a simple univariate linear model:
\begin{flalign}\label{linear_model}
Z_{1t}=\alpha_0+\beta_0 Z_{2t}+\varepsilon_t,
\end{flalign}
where $Z_t=(Z_{1t},Z_{2t})^{\top}$, $\ct_0=(\alpha_0,\beta_0)^{\top}$, and $E(\varepsilon_t|X_t)=0$.
Model (\ref{linear_model}) is corresponding to model (\ref{model}) with $h(Z_t,\ct_0)=Z_{1t}-\alpha_0-\beta_0 Z_{2t}$.
Let $\ct=(\alpha, \beta)^{\top}$ and $h(Z_t,\ct)=Z_{1t}-\alpha-\beta Z_{2t}$. For this model,
it is straightforward to see
\begin{flalign*}
E\big[h(Z_t,\ct)-E(h(Z_t,\ct))|X_t\big]&=(\beta_0-\beta)E[Z_{2t}-E(Z_{2t})|X_t],\\
E(h(Z_t,\ct)|X_t)&=(\alpha_0-\alpha)+(\beta_0-\beta)E(Z_{2t}|X_t).
\end{flalign*}
Therefore, under some mild conditions on $X_t$ and $Z_{2t}$, the MDD-based estimator $\widehat{\theta}_n$ can only identify $\beta_0$, while
the estimator in \citet{DL:2004} can identify both $\alpha_0$ and $\beta_0$.
This simple example reveals that
the identification condition in Assumption \ref{hatii} only allows us to deal with a model without intercept parameter.
 For the model with intercept parameter, a two-step estimation procedure can be given to estimate both non-intercept and intercept parameters
(see more discussions for model (\ref{special_model}) below).

Define
\begin{flalign}\label{def_u}
u(x)=E\Big\{\Big[\frac{\partial h(Z_t,\ct_0)}{\partial \theta}-E\Big(\frac{\partial h(Z_t,\ct_0)}{\partial \theta}\Big)\Big]\|X_t-x\|\Big\}.
\end{flalign}
Below, we show the consistency and asymptotic normality of $\widehat{\theta}_n$.

\begin{thm}\label{thm1}
Suppose Assumptions \ref{hatii}--\ref{interpoint} hold. Then,
$\widehat{\theta}_n-\theta_0\overset{p}{\longrightarrow}0$
as $n\to\infty$.
\end{thm}

\begin{thm}\label{thm2}
Suppose Assumptions \ref{hatii}--\ref{whitenoise} hold. Then,
\begin{equation*}
\sqrt{n}(\widehat{\theta}_n-\ct_0)\overset{d}{\longrightarrow} N(0,\W^{-1}\S\W^{-1}) \,\,\,\mbox{ as }n\to\infty,
\end{equation*}
where
\begin{flalign*}
\W&=E\Big\{\Big[\frac{\partial h(Z_t,\ct_0)}{\partial \theta}-E\Big(\frac{\partial h(Z_t,\ct_0)}{\partial \theta}\Big)\Big]^\top  u(X_t)\Big\},\\
\S&=E\big\{[u(X_t)-E(u(X_t))]^\top h(Z_t,\ct_0)h(Z_t,\ct_0)^\top [u(X_t)-E(u(X_t))]\big\}.
\end{flalign*}
\end{thm}

To establish the asymptotic properties of $\widehat{\theta}_n$, our objective is to transform $\mbox{MDD}_n(\theta)$ into an integral form.
For this purpose, we need an identity
\begin{flalign}\label{identity}
	\|x\|=\int_{\mathbb{R}^{q}}\big[1-\cos(\langle s,x\rangle)\big]w_{*}(s)ds \,\,\mbox{ for all }x\in \mathbb{R}^{q};
\end{flalign}
see \citet{SRB:2007}. Using the identity in (\ref{identity}) and the fact that
\begin{flalign}\label{key_result}
	\sum_{t=1}^{n}\sum_{t'=1}^{n}
	\Big\{\big[\big(h(Z_t,\theta)-\bar{h}(\theta)\big)^{\top}\big(h(Z_{t'},\theta)-\bar{h}(\theta)\big)\big]\Big\}=0,
\end{flalign}
we can show
\begin{flalign*}
	\begin{split}
		\mbox{MDD}_n(\theta)&=\frac{1}{n^2}
		\sum_{t=1}^{n}\sum_{t'=1}^{n}
		\Big\{\big[\big(h(Z_t,\theta)-\bar{h}(\theta)\big)^{\top}\big(h(Z_{t'},\theta)-\bar{h}(\theta)\big)\big]\\
		&\quad\quad\quad\quad \times \int_{\mathbb{R}^{q}}\cos(\langle s,X_{t'}-X_{t}\rangle)w_{*}(s)ds\Big\}\\
		&=\frac{1}{n^2}
		\sum_{t=1}^{n}\sum_{t'=1}^{n}
		\Big\{\big[\big(h(Z_t,\theta)-\bar{h}(\theta)\big)^{\top}\big(h(Z_{t'},\theta)-\bar{h}(\theta)\big)\big]\\
		&\quad\quad\quad\quad \times \int_{\mathbb{R}^{q}}\big[\cos(\langle s,X_{t'}-X_{t}\rangle)+\textrm{i} \sin(\langle s,X_{t'}-X_{t}\rangle)\big]w_{*}(s)ds\Big\},
	\end{split}
\end{flalign*}
where the last result holds because $\sin(\cdot)$ is an odd function. By observing that
$\cos(\langle s,X_{t'}-X_{t}\rangle)+\textrm{i} \sin(\langle s,X_{t'}-X_{t}\rangle)=\exp(\textrm{i}\langle s,X_{t'}\rangle)
\exp(-\textrm{i}\langle s,X_{t}\rangle)$, it entails
\begin{flalign}\label{mdd_sample_integral}
	\mbox{MDD}_n(\theta)&=
	\int_{\mathbb{R}^{q}} \Big|\mathcal{G}_n(s,\ct)\Big|_{l}^{2}w_*(s) ds
	=\int_{\mathbb{R}^{q}} \mathcal{G}_n(s,\ct)^{\star} \mathcal{G}_n(s,\ct)w_*(s) ds,
\end{flalign}
where
\begin{equation}\label{g_n_s_theta}
	\mathcal{G}_n(s,\ct)=\frac{1}{n}\sum_{t=1}^{n} \big(h(Z_t,\ct)-\bar{h}(\theta)\big)\exp(\textrm{i}\langle s,X_t\rangle).
\end{equation}
The integral form of $\mbox{MDD}_n(\theta)$ in (\ref{mdd_sample_integral}) enables us to investigate the asymptotics of $\widehat{\theta}_n$ by studying those of the process $\mathcal{G}_n(s,\ct)$. This feature makes our proof techniques applicable to time series samples, so it largely
broadens the application scope of $\widehat{\theta}_n$ for studying the multiple linear/nonlinear time series models.

It is worth noting that our proof techniques can not apply to the ICM estimator in \citet{Tsyawo:2023}. This is because the sample counterpart of
$\mbox{ICM}(\theta_0)$ in (\ref{eqn1_8}) is
\begin{flalign*}
	\mbox{ICM}_n(\theta)&=-\frac{1}{n^2}
	\sum_{t=1}^{n}\sum_{t'=1}^{n}
	\Big\{\big[h(Z_t,\theta)^{\top}h(Z_{t'},\theta)\big]\|X_t-X_{t'}\|\Big\},
\end{flalign*}
based on a similar argument as for (\ref{mdd_samle_form}). However, it is not possible to transform $\mbox{ICM}_n(\theta)$ into a similar integral form as $\mbox{MDD}_n(\theta)$, since the result (\ref{key_result}) does not hold in the absence of $\bar{h}(\theta)$.

To see the linkage between our asymptotic normality result and that in \citet{Tsyawo:2023}, we consider the following multiple linear model (without a constant vector):
\begin{flalign}\label{linear_model}
Z_{1t}=\Gamma_0 Z_{2t}+\varepsilon_t,
\end{flalign}
where $Z_{1t}\in \mathbb{R}^{l}$, $Z_{2t}\in\mathbb{R}^{(k-l)}$, $\Gamma_0\in\mathbb{R}^{l\times (k-l)}$, and $E(\varepsilon_t|X_t)=0$.
Then, model (\ref{linear_model}) is a special case of model (\ref{model}) with $Z_t=(Z_t^{\top},Z_{2t}^{\top})^{\top}$, $\theta_0=vec(\Gamma_0)$, and $h(Z_t,\theta_0)=Z_{1t}-\Gamma_0 Z_{2t}$.
In this case, it is not hard to see that $\widehat{\theta}_n=vec(\widehat{\Gamma}_{n})$ has a closed-form solution:
$$\widehat{\theta}_n=\Xi_{1n}^{-1}\Xi_{2n},$$
where $$\Xi_{1n}=\sum_{t=1}^n\sum_{t'=1}^n\Big[\mbox{I}_{l}\otimes Z_{2t}^{\top}-\frac{1}{n}\sum_{t=1}^n\big(\mbox{I}_{l}\otimes Z_{2t}^{\top}\big)\Big]^\top\Big[\mbox{I}_{l}\otimes Z_{2t'}^{\top}-\frac{1}{n}\sum_{t=1}^n\big(\mbox{I}_{l}\otimes Z_{2t}^{\top}\big)\Big]\left\|Z_{2t}-Z_{2t'}\right\|$$
and
$$\Xi_{2n}=\sum_{t=1}^n\sum_{t'=1}^n\Big[\mbox{I}_{l}\otimes Z_{2t}^{\top}-\frac{1}{n}\sum_{t=1}^n\big(\mbox{I}_{l}\otimes Z_{2t}^{\top}\big)\Big]^\top \Big(Z_{1t'}-\frac{1}{n}\sum_{t=1}^n Z_{1t}\Big)\left\|Z_{2t}-Z_{2t'}\right\|. $$
In addition, for model (\ref{linear_model}), we have $\partial h(Z_t,\ct_0)/\partial \theta=-\mbox{I}_{l}\otimes Z_{2t}^{\top}$, leading to
\begin{flalign*}
\W&=E\big\{\big[\mbox{I}_{l}\otimes Z_{2t}^{\top}-E\big(\mbox{I}_{l}\otimes Z_{2t}^{\top}\big)\big]^\top  u(X_t)\big\},\\
\S&=E\big\{[u(X_t)-E(u(X_t))]^\top \varepsilon_t\varepsilon_t^\top [u(X_t)-E(u(X_t))]\big\},
\end{flalign*}
where $u(x)=E\big\{\big[\mbox{I}_{l}\otimes Z_{2t}^{\top}-E\big(\mbox{I}_{l}\otimes Z_{2t}^{\top}\big)\big]\|X_t-x\|\big\}$.  In other words, if $E(u(X_t))=0$, $E(Z_{2t})=0$, and $l=1$, our asymptotic variance
$\W^{-1}\S\W^{-1}$ becomes that of \citet{Tsyawo:2023}.

\section{The Two-step Estimation Procedure}\label{sec_3}

In this section, we provide a two-step estimation procedure for model (\ref{model}) with intercept parameters. Specifically,  we consider a special model (\ref{model}) with
\begin{flalign}\label{special_model}
h(Z_t,\theta_0)=\Big(
\begin{array}{c}
\theta_{10}\\
0
\end{array}\Big)
+m(Z_t,\theta_{20}),
\end{flalign}
where $\theta_0=(\theta_{10}^{\top},\theta_{20}^{\top})^{\top}$,
$\theta_{10}$ is a $d_1$-dimensional vector of intercept parameters with $d_1\leq l$, $\theta_{20}$ is a $d_2$-dimensional vector of non-intercept parameters,
and $m$ is a given function mapping $\mR^k\times\CT$ into $\mR^l$.
Clearly, model (\ref{special_model}) nests model (\ref{linear_model}) with $d_1=l=1$, $d_2=1$, $\theta_{10}=-\alpha_0$, $\theta_{20}=\beta_0$, and $m(Z_t,\theta_{20})=Z_{1t}-\theta_{20}Z_{2t}$.
Let $\theta=(\theta_{1}^{\top},\theta_{2}^{\top})^{\top}\in \Theta_1\times \Theta_2$, where $\Theta_i\subset \mathbb{R}^{d_i}$ for $i=1$ and 2.
As discussed before, the MDD-based estimation method can not identify $\theta_{10}$, so it has to exclude the estimation of $\theta_{10}$ by simply replacing $h(Z_t,\ct)$ with $m(Z_t,\theta_{2})$.
Since the value of $\mbox{MDD}_n(\theta)$ in (\ref{mdd_samle_form}) is invariant to any shift on $h(Z_t,\ct)$, this
exclusion implementation does not impact the MDD-based estimator of $\theta_{20}$ given by
\begin{flalign*} 
\widehat{\theta}_{2n}=\mathop{\arg\min}\limits_{\ct_2 \in \CT_2}\operatorname{MDD}_{2,n}(\theta_2),
\end{flalign*}
where $\operatorname{MDD}_{2,n}(\theta_2)$ is defined in the same way as $\operatorname{MDD}_{n}(\theta)$, with
$h(Z_t,\theta)$ replaced by $m(Z_t,\theta_2)$.
After obtaining $\widehat{\theta}_{2n}$, we partition the functions $h$ and $m$ as follows:
\begin{flalign}\label{partition}
h(Z_t,\theta_0)=
\Big(
\begin{array}{c}
h_1(Z_t,\theta_0)\\
h_2(Z_t,\theta_0)
\end{array}\Big)\,\,\,
\mbox{ and }\,\,\,
m(Z_t,\theta_{20})=
\Big(
\begin{array}{c}
m_1(Z_t,\theta_{20})\\
m_2(Z_t,\theta_{20})
\end{array}\Big),
\end{flalign}
with $h_1\in \mathcal{R}^{d_1}$, $h_2\in \mathcal{R}^{l-d_1}$, $m_1\in \mathcal{R}^{d_1}$, and $m_2\in \mathcal{R}^{l-d_1}$.
Then, we use the estimating moment functions $E(\theta_{10}+m_1(Z_t,\theta_{20}))=0$
to estimate $\theta_{10}$ by
\begin{flalign*} 
\widehat{\theta}_{1n}=-\frac{1}{n}\sum_{t=1}^{n} m_1(Z_t,\widehat{\theta}_{2n}).
\end{flalign*}

With a slight abuse of notation, we denote $\widehat{\theta}_n=(\widehat{\theta}_{1n}^{\top}, \widehat{\theta}_{2n}^{\top})^{\top}$ for model (\ref{special_model}).
To present the asymptotic results of $\widehat{\theta}_{1n}$ and $\widehat{\theta}_{2n}$, we need the following notation:
\begin{flalign}\label{def_u_2}
\begin{split}
u_2(x)&=E\Big\{\Big[\frac{\partial m(Z_t,\ct_{20})}{\partial \theta_2}-E\Big(\frac{\partial m(Z_t,\ct_{20})}{\partial \theta_2}\Big)\Big]\|X_t-x\|\Big\},\\
\W_1&=E\big\{h(Z_t,\ct_0)h(Z_t,\ct_0)^\top [u_2(X_t)-E(u_2(X_t))]\big\},\\
\S_1&=E\big\{h(Z_t,\ct_0)h(Z_t,\ct_0)^\top\big\},\\
\W_2&=E\Big\{\Big[\frac{\partial m(Z_t,\ct_{20})}{\partial \theta_2}-E\Big(\frac{\partial m(Z_t,\ct_{20})}{\partial \theta_2}\Big)\Big]^\top  u_2(X_t)\Big\},\\
\S_2&=E\big\{[u_2(X_t)-E(u_2(X_t))]^\top h(Z_t,\ct_0)h(Z_t,\ct_0)^\top [u_2(X_t)-E(u_2(X_t))]\big\},\\
J_t&=
\begin{pmatrix}
E\Big(\frac{\partial m_1(Z_t,\theta_{20})}{\partial\theta_2}\Big) \Omega_2^{-1} [u_2(X_t)-E(u_2(X_t))]^{\top}-\Upsilon\\
-\Omega_2^{-1} [u_2(X_t)-E(u_2(X_t))]^{\top}
\end{pmatrix}
\end{split}
\end{flalign}
with $\Upsilon=(\mbox{I}_{d_1},0)\in\mathcal{R}^{d_1\times l}$. Here, $\mbox{I}_{d_1}$ is the identity matrix of size $d_1$.
Under an additional assumption for the identifiability of $\theta_{20}$, we are able to establish the joint asymptotic normality of
 $\widehat{\theta}_{1n}$ and $\widehat{\theta}_{2n}$.

\begin{assum}
\label{hatii_special}
$E\big[m(Z_t,\ct_2)-E(m(Z_t,\ct_2))|X_t\big]=0$ (a.s.) if and only if $\ct_2=\ct_{20}$.
\end{assum}

\begin{thm}\label{thm3}
Suppose Assumptions \ref{stationarity}--\ref{whitenoise} and \ref{hatii_special} hold. Then, under model (\ref{special_model}),
\begin{equation*}
\sqrt{n}(\widehat{\theta}_n-\theta_0)\overset{d}{\longrightarrow} N(0,V) \,\,\,
\mbox{ as }n\to\infty,
\end{equation*}
where $V=E\big[J_t h(Z_t,\ct_0)h(Z_t,\ct_0)^\top J_t^{\top}\big]$. Particularly,
\begin{equation*}
\sqrt{n}(\widehat{\theta}_{1n}-\ct_{10})\overset{d}{\longrightarrow} N(0,V_1)
\mbox{ and } \sqrt{n}(\widehat{\theta}_{2n}-\ct_{20})\overset{d}{\longrightarrow} N(0,V_2)\,\,\,\mbox{ as }n\to\infty,
\end{equation*}
where
\begin{flalign*}
V_1&=E\Big(\frac{\partial m_1(Z_t,\theta_{20})}{\partial\theta_2}\Big) \Omega_2^{-1}\Sigma_2\Omega_2^{-1}E\Big(\frac{\partial m_1(Z_t,\theta_{20})}{\partial\theta_2}\Big)^{\top}-\Upsilon\Omega_1\Omega_2^{-1}E\Big(\frac{\partial m_1(Z_t,\theta_{20})}{\partial\theta_2}\Big)^{\top}\\
&\quad-E\Big(\frac{\partial m_1(Z_t,\theta_{20})}{\partial\theta_2}\Big)\Omega_2^{-1}\Omega_1^{\top}\Upsilon^{\top}
+\Upsilon\Sigma_1\Upsilon^{\top},\\
V_2&=\W_2^{-1}\S_2\W_2^{-1}.
\end{flalign*}
\end{thm}

As the asymptotic variances in Theorems \ref{thm2} and \ref{thm3} have analytic expressions, they can be directly estimated by their sample counterparts. This leads to an easy-to-implement statistical inference for $\theta_0$, without the selection of any tuning parameters and preliminary estimator. Meanwhile, we should highlight that our technical assumptions for both theorems allow the data to have the conditional heteroskedasticity of unknown form, so it makes our estimation method have a large application scope to handle the multiple linear/nonlinear time series models.

Finally, we give some remarks on the estimation efficiency. First, although it is hard to make a formal efficiency comparison between $\widehat{\theta}_n$ and the estimator in \citet{DL:2004}, the simulation studies in the next section demonstrate that $\widehat{\theta}_n$ is significantly more efficient than the estimator in \citet{DL:2004}. Second,
$\widehat{\theta}_n$ is not granted to be efficient.
Following \citet{DL:2004}, we can take $\widehat{\theta}_n$ as the preliminary estimator to obtain an efficient estimator
$$\widetilde{\theta}_n=\widehat{\theta}_n-\Big[\frac{\partial^2 L_n(\widehat{\theta}_n)}{\partial\theta\partial\theta^{\top}}\Big]^{-1}\frac{\partial L_n(\widehat{\theta}_n)}{\partial \theta},$$
where $L_n(\theta)$ is the efficient objective function. However, the derivation of $L_n(\theta)$ usually relies on some nonparametric methods to obtain optimal IVs and so the selection of tuning parameters is unavoidable; see \citet{Newey:1993}. Since our goal is to provide a user-friendly statistical inference for $\theta_0$, the study of $\widetilde{\theta}_n$ will not be given in this paper and is left for future research.

\section{Simulations}\label{sec_4}
In this section, we carry out simulation experiments to assess the finite-sample performance of our proposed MDD-based estimator $\widehat{\theta}_n$. First, we generate $1000$ Monte-Carlo repetitions of sample size $n=50$, $100$, and $200$ from the following twelve univariate data generating processes (DGPs):

\begin{itemize}[label={}]
\item DGP 1: $Z_{1t}=\theta_0Z_{2t}+\varepsilon_t$, where $\theta_0=1$, $Z_{2t}$ satisfies the order 1 autoregressive (AR(1)) model: $Z_{2t}=0.3Z_{2,t-1}+\eta_t$, and $(\varepsilon_t,\eta_t)^{\top}\overset{i.i.d.}{\sim} N(0,\mbox{I}_2)$;

\item DGP 2: $Z_{1t}=\theta_0Z_{2t}+\varepsilon_t$, where $\theta_0=1$, $Z_{2t}$ satisfies the AR(1) model: $Z_{2t}=0.3Z_{2,t-1}+\zeta_t$, $\varepsilon_t$ satisfies the order 1 autoregressive conditional heteroskedasticity (ARCH(1)) model: $\varepsilon_t=v_t^{1/2}\eta_t$ and $v_t=0.4+0.5\varepsilon_{t-1}^2$, and $(\zeta_t,\eta_t)^{\top}\overset{i.i.d.}{\sim} N(0,\mbox{I}_2)$;

\item DGP 3: $Z_{1t}=\sin(\theta_0Z_{2t})+\varepsilon_t$, where $\theta_0=1$, $\varepsilon_t\overset{i.i.d.}{\sim} N(0,1)$, $Z_{2t}\overset{i.i.d.}{\sim} U[-1,1]$, and $\varepsilon_t$ and $Z_{2t}$ are independent;

\item DGP 4: $Z_{1t}=\sin(\theta_0Z_{2t})+\varepsilon_t$, where $\theta_0=1$, $\varepsilon_t$ follows the ARCH(1) model: $\varepsilon_t=v_t^{1/2}\eta_t$ and $v_t=0.4+0.5\varepsilon_{t-1}^2$, $\eta_t\overset{i.i.d.}{\sim} N(0,1)$, $Z_{2t}\overset{i.i.d.}{\sim} U[-1,1]$, and $\eta_t$ and  $Z_{2t}$ are independent;

\item DGP 5: $Z_{1t}=\text{sigmoid}(\theta_0Z_{2t})+\varepsilon_t$, where $\theta_0=1$, $\text{sigmoid}(x)=1/[1+\exp(-x)]$,
and $\varepsilon_t$ and $Z_{2t}$ are generated as in DGP 3;


\item DGP 6: $Z_{1t}=\text{sigmoid}(\theta_0Z_{2t})+\varepsilon_t$, where $\theta_0=1$, and $\varepsilon_t$ and $Z_{2t}$ are generated as in DGP 4;


\item DGP 7: $Z_{1t}=\ct_0^2Z_{2t}+\ct_0Z_{2t}^2+\varepsilon_t$, where $\ct_0=5/4$ and $(\epsilon_t,Z_{2t})^{\top}\overset{i.i.d.}{\sim} N(0,\mbox{I}_2)$;

\item DGP 8: $Z_{1t}=\theta_0Z_{2t}+\varepsilon_t$, where $\theta_0=1$, $\varepsilon_t$ satisfies the AR(1) model: $\varepsilon_t=0.1\varepsilon_{t-1}+\eta_t$, and $(\eta_t,Z_{2t})^{\top}\overset{i.i.d.}{\sim} N(0,\mbox{I}_2)$;

\item DGP 9: $Z_{1t}=\theta_0Z_{1,t-1}+\varepsilon_t$, where $\theta_0=0.5$ and $\varepsilon_t \overset{i.i.d.}{\sim} t(7)$;

\item DGP 10: $Z_{1t}=\theta_0Z_{1,t-1}+\varepsilon_t$, where $\theta_0=0.5$ and $\varepsilon_t$ is generated as in DGP 2;



\item DGP 11: $Z_{1t}=\theta_{10}+\theta_{20}Z_{2t}+\varepsilon_t$, where $\theta_0=(\theta_{10},\theta_{20})^{\top}=(0.5, 1)^{\top}$, and $\varepsilon_t$ and $Z_{2t}$ are generated as in DGP 1;



\item DGP 12: $Z_{1t}=\theta_{10}+\theta_{20}Z_{2t}+\varepsilon_t$, where $\theta_0=(\theta_{10},\theta_{20})^{\top}=(0.5, 1)^{\top}$,
$Z_{2t}=\zeta_t$, and $\varepsilon_t$ and $\zeta_t$ are generated as in DGP 2.

\end{itemize}

\noindent
DGPs 1--10 consider ten models without intercept, while DPGs 11--12 regard two models with intercept.
Among them, DGPs 1--8 and 11--12 are
linear/nonlinear time series models or regressions with i.i.d. errors (DGPs 1, 3, 5, 7, and 11), conditionally heteroskedastic errors (DGPs 2, 4, 6, and 12), or correlated errors (DGP 8), and DGPs 9 and 10 are AR models with i.i.d. errors and conditionally heteroskedastic errors, respectively.
We compute the MDD-based estimator $\widehat{\theta}_n$ of $\theta_0$ in DGPs 1--10, while we
calculate $\widehat{\theta}_n=(\widehat{\theta}_{1n},\widehat{\theta}_{2n})^{\top}$ of $\theta_0$ in DGPs 11--12 by using the two-step estimation procedure. As a comparison, we also compute the estimator in \citet{DL:2004} for each DGP. For all considered estimators above, we use the conditioning variable
$Z_{1,t-1}$ for DGPs 9--10 and $Z_{2t}$ for the other DGPs.

Based on the results from $1000$ repetitions, Table \ref{table1} reports the bias, empirical standard deviation (ESD), and asymptotic standard deviation (ASD) of all considered estimators, where our proposed estimators and the estimator in \citet{DL:2004} are abbreviated as ``MDD'' and ``DL'', respectively. From this table, we find that except for DGPs 5--6, both MDD and DL estimators have small biases, and the value of ASD is close to that of ESD for both estimators, in align with their related asymptotic normality results.
 For DGPs 5--6, both MDD and DL estimators have large biased when $n=50$ and $100$, and the bias of DL estimator remains large when $n=200$;
 in these two DGPs, the values of ASD and ESD for the MDD estimator get closed as $n$ increases to 200, whereas those for the DL estimator still have a significant difference for a large $n$ particularly when the errors are conditionally heteroskedastic.
 Note that our findings in DGPs 5--6 match the statement on page 163 of \cite{Tsay:2005} that the parameters in the sigmoid or smooth transition function are hard to estimate. Nevertheless, our simulations results in DGPs 5--6 show that the MDD estimator can outperform DL estimator significantly for the smooth transition model, especially when the sample size is small. Moreover, from Table \ref{table1}, we observe that
 the MDD estimator always has a much smaller value of ASD than the DL estimator, indicating a significant efficiency advantage of MDD estimator over DL estimator in all examined DGPs.


Next, we generate $1000$ Monte-Carlo repetitions of sample size $n=50$, $100$, and $200$ from the following four multivariate DGPs:
\begin{itemize}[label={}]
\item DGP 13: $Z_{1t}=A_0 Z_{2t}+\varepsilon_t$, where
$A_0=\begin{pmatrix}
    \theta_{11,0}& \theta_{12,0}\\ \theta_{21,0}& \theta_{22,0}\end{pmatrix}$ with $\theta_0=(\theta_{11,0},\theta_{12,0},\theta_{21,0},\theta_{22,0})^{\top}$ $=(1,-1,1,2)^{\top}$, $Z_{2t}=\begin{pmatrix}
    0.3 & 0\\
    0 & 0.2
    \end{pmatrix}Z_{2,t-1}+\zeta_t$, and $(\zeta_t^{\top},\varepsilon_t^{\top})^{\top}\overset{i.i.d.}{\sim} N(0, \mbox{I}_{4})$;

\item DGP 14: $Z_{1t}=A_0 Z_{2t}+\varepsilon_t$, where $A_0$ and $Z_{2t}$ are defined as in DGP 13,  $\varepsilon_t=(\varepsilon_{1,t},\varepsilon_{2,t})^{\top}$ follows the multivariate generalized ARCH (GARCH) model:
    $\varepsilon_t=V_t^{1/2}\eta_t$ and $V_t=\left(v_{i j,t}\right)_{i, j=1,2}$ is a $2\times 2$ symmetric matrix with
$$
\left\{\begin{array}{l}
v_{11,t}=0.1+0.8 v_{11,t-1}+0.1 \varepsilon_{1,t-1}^2, \\
v_{22,t}=0.1+0.8 v_{22,t-1}+0.1 \varepsilon_{2,t-1}^2, \\
v_{12,t}=0.7 \sqrt{v_{11,t} v_{22,t}},
\end{array}\right.
$$
and $(\zeta_t^{\top},\eta_t^{\top})^{\top}\overset{i.i.d.}{\sim} N(0, \mbox{I}_{4})$;

\item DGP 15: $Z_{1t}=A_0 Z_{2t}+\varepsilon_t$, where $A_0$ and $Z_{2t}$ are defined as in DGP 13, $\varepsilon_t$ follows the vector AR(1) model:
$\varepsilon_t=\begin{pmatrix}
    0.2 & 0\\
    0 & 0.1
    \end{pmatrix} \varepsilon_{t-1}+\eta_t$, and $(\zeta_t^{\top},\eta_{t}^{\top})^{\top}\overset{i.i.d.}{\sim} N(0, \mbox{I}_{4})$;

\item DGP 16: $Z_{1t}=A_0 Z_{1,t-1}+\varepsilon_t$, where $A_0$ is defined as in DGP 13 with
$\theta_0=(0.6,-0.4,0.8,0.2)^{\top}$, and $\varepsilon_t\overset{i.i.d.}{\sim} N(0, \mbox{I}_{2})$.
\end{itemize}

\noindent DGPs 13, 14, and 15 are three time series models with i.i.d. errors, conditionally heteroskedastic errors, and correlated errors, respectively, and DGP 16 is a vector AR(1) model with i.i.d. errors. For all of these DGPs, we compute the MDD estimator $\widehat{\theta}_n$
with the conditioning variable $Z_{2t}$ for DGPs 13--15 and $Z_{1,t-1}$ for DGP 16. As a comparison, the DL estimator is also
calculated with the same choices of conditioning variable as the MDD estimator.

Based on the results from $1000$ repetitions, Table \ref{table2} reports the bias, ESD, and ASD of both MDD and DL estimators
in DGPs 13--16. From this table, we can reach the similar conclusion as above that
the MDD estimators are more efficient than the DL estimators in all examined DGPs.

\begin{table}[!ht]
\caption{The estimation results for DGPs 1--12}
\label{table1}
   \setlength{\tabcolsep}{1.4mm}{
\begin{tabular}{lccccccccccccc}
\toprule
                      &     & &\multicolumn{3}{c}{$n=50$} &  & \multicolumn{3}{c}{$n=100$} &  & \multicolumn{3}{c}{$n=200$} \\
                      \cmidrule(lr){4-6} \cmidrule(lr){8-10} \cmidrule(lr){12-14}
                      &     & & Bias    & ASD    & ESD   &  & Bias   & ASD    & ESD    &  & Bias    & ASD    & ESD    \\
                      \cmidrule(lr){1-14}
\multirow{2}{*}{DGP 1} & MDD & $\theta_0$ &-0.007  & 0.143  & 0.148 &  & 0.001   & 0.098  & 0.098  &  & -0.002  & 0.069  & 0.072  \\
                      & DL & $\theta_0$ &-0.019  & 0.238  & 0.244 &  & -0.003  & 0.175  & 0.173  &  & -0.003  & 0.125  & 0.129  \\
                      &     &         &        &       &  &         &        &        &  &         &        &        \\
\multirow{2}{*}{DGP 2} & MDD & $\theta_0$ &-0.001   & 0.120  & 0.127 &  & -0.001   & 0.086  & 0.087  &  & 0.001   & 0.061  & 0.059  \\
                      & DL & $\theta_0$ &-0.004   & 0.207  & 0.220 &  & -0.003   & 0.154  & 0.161  &  & -0.000    & 0.111  & 0.110  \\
                      &     &         &        &       &  &         &        &        &  &         &        &        \\
\multirow{2}{*}{DGP 3} & MDD & $\theta_0$ &0.063   & 0.400  & 0.389 &  & 0.011   & 0.254  & 0.256  &  & 0.080   & 0.174  & 0.175  \\
                      & DL & $\theta_0$ &0.060   & 0.668  & 0.556 &  & 0.015   & 0.458  & 0.422  &  & 0.017   & 0.311  & 0.296   \\
                      &     &         &        &       &  &         &        &        &  &         &        &        \\
\multirow{2}{*}{DGP 4} & MDD & $\theta_0$ &0.046   & 0.374  & 0.336 &  & 0.006   & 0.229  & 0.229  &  & 0.005   & 0.155  & 0.154  \\
                      & DL & $\theta_0$ &0.055   & 0.558  & 0.514 &  & 0.013   & 0.390  & 0.376  &  & 0.014   & 0.270  & 0.261  \\
                      &     &         &        &       &  &         &        &        &  &         &        &        \\
\multirow{2}{*}{DGP 5} & MDD & $\theta_0$ &0.647  & 7.023  & 5.902 &  & 0.279  & 3.226  & 3.008  &  & 0.060  & 0.465  & 0.507  \\
                      & DL & $\theta_0$ &1.861  & 40.682  & 10.058 &  & 0.922  & 13.229  & 6.025  &  & 0.239  & 2.432  & 1.883  \\
                      &     &         &        &       &  &         &        &        &  &         &        &        \\
\multirow{2}{*}{DGP 6} & MDD & $\theta_0$ &0.884  & 7.933  & 6.414 &  & 0.264  & 1.145  & 1.262  &  & 0.078  & 0.550  & 0.596  \\
                      & DL & $\theta_0$ &2.308  & 63.365  & 12.546 &  & 1.071  & 13.743  & 6.426  &  & 0.310  & 2.436  & 2.029  \\
                      &     &         &        &       &  &         &        &        &  &         &        &        \\
\multirow{2}{*}{DGP 7} & MDD & $\theta_0$ &-0.005  & 0.060  & 0.059 &  & 0.000   & 0.041  & 0.039  &  & -0.001  & 0.029  & 0.028  \\
                      & DL & $\theta_0$ &-0.007  & 0.110  & 0.117 &  & 0.001   & 0.077  & 0.079  &  & -0.000  & 0.056  & 0.057  \\
                      &     &         &        &       &  &         &        &        &  &         &        &        \\
\multirow{2}{*}{DGP 8} & MDD & $\theta_0$ &-0.010  & 0.148  & 0.152 &  & 0.000  & 0.103  & 0.103  &  & -0.002  & 0.073  & 0.074  \\
                      & DL & $\theta_0$ &-0.025  & 0.258  & 0.277 &  & -0.004  & 0.186  & 0.196  &  & -0.003  & 0.131  & 0.145  \\
                      &     &         &        &       &  &         &        &        &  &         &        &        \\
\multirow{2}{*}{DGP 9} & MDD & $\theta_0$ &-0.043  & 0.130  & 0.130 &  & -0.023  & 0.091  & 0.093  &  & -0.012  & 0.064  & 0.063  \\
                      & DL & $\theta_0$ &-0.009  & 0.216  & 0.209 &  & -0.024  & 0.161  & 0.154  &  & -0.022  & 0.116  & 0.113  \\
                      &     &         &        &       &  &         &        &        &  &         &        &        \\
\multirow{2}{*}{DGP 10} & MDD & $\theta_0$ &-0.048  & 0.153  & 0.155 &  & -0.016  & 0.110  & 0.114  &  & -0.004  & 0.079  & 0.083  \\
                      & DL & $\theta_0$ &-0.012  & 0.229  & 0.220 &  & -0.026  & 0.171  & 0.163  &  & -0.018  & 0.125  & 0.121  \\
                      &     &         &        &       &  &         &        &        &  &         &        &        \\
\multirow{4}{*}{DGP 11} & \multirow{2}{*}{MDD} & $\theta_{10}$ & -0.007  & 0.139  & 0.148 &  & 0.001   & 0.097  & 0.098  &  & -0.002  & 0.069  & 0.072  \\
                      &                      & $\theta_{20}$ & 0.008   & 0.142  & 0.141 &  & 0.003   & 0.100  & 0.099  &  & 0.000   & 0.070  & 0.071  \\
                      & \multirow{2}{*}{DL}  & $\theta_{10}$ & -0.008  & 0.163  & 0.170 &  & 0.000   & 0.115  & 0.115  &  & -0.004  & 0.081  & 0.086  \\
                      &                      & $\theta_{20}$ & 0.008   & 0.149  & 0.147 &  & 0.002   & 0.105  & 0.105  &  & -0.001  & 0.074  & 0.076  \\
                      &                      &       &         &        &       &  &         &        &        &  &         &        &        \\
\multirow{4}{*}{DGP 12} & \multirow{2}{*}{MDD} & $\theta_{10}$ & -0.007  & 0.125  & 0.133 &  & 0.000   & 0.089  & 0.091  &  & -0.002  & 0.064  & 0.066  \\
                      &                      & $\theta_{20}$ & 0.006   & 0.120  & 0.125 &  & 0.001   & 0.087  & 0.088  &  & 0.000   & 0.062  & 0.063  \\
                      & \multirow{2}{*}{DL}  & $\theta_{10}$ & -0.009  & 0.145  & 0.158 &  & -0.000  & 0.104  & 0.108  &  & -0.003  & 0.075  & 0.078  \\
                      &                      & $\theta_{20}$ & 0.005   & 0.126  & 0.131 &  & 0.001   & 0.091  & 0.093  &  & -0.001  & 0.066  & 0.068  \\
                      \bottomrule
\end{tabular}
}
\end{table}

\begin{table}[!ht]
\caption{The estimation results for DGPs 13--16}\label{table2}
\centering
\setlength{\tabcolsep}{1.5mm}
\begin{tabular}{cccccccccccccc}
\toprule
                      &                      &       & \multicolumn{3}{c}{$n=50$} &  & \multicolumn{3}{c}{$n=100$} &  & \multicolumn{3}{c}{$n=200$} \\ \cmidrule(lr){4-6} \cmidrule(lr){8-10} \cmidrule(lr){12-14}
                      &                      &       & BIAS    & ASD    & ESD   &  & BIAS    & ASD    & ESD    &  & BIAS    & ASD    & ESD    \\ \cmidrule(lr){1-14}
\multirow{8}{*}{DGP 13}  & \multirow{4}{*}{MDD} & $\theta_{11,0}$ & 0.001   & 0.138  & 0.146 &  & 0.002   & 0.097  & 0.099  &  & 0.001   & 0.069  & 0.071  \\
                       &                      & $\theta_{12,0}$ & -0.002  & 0.139  & 0.141 &  & -0.001  & 0.097  & 0.100  &  & -0.001  & 0.069  & 0.069  \\
                       &                      & $\theta_{21,0}$ & 0.002   & 0.141  & 0.148 &  & 0.002   & 0.100  & 0.106  &  & 0.005   & 0.070  & 0.070  \\
                       &                      & $\theta_{22,0}$ & 0.001   & 0.142  & 0.152 &  & -0.004  & 0.100  & 0.104  &  & -0.000  & 0.070  & 0.070  \\
                       & \multirow{4}{*}{DL}  & $\theta_{11,0}$ & 0.003   & 0.206  & 0.219 &  & 0.001   & 0.145  & 0.151  &  & 0.002   & 0.102  & 0.107  \\
                       &                      & $\theta_{12,0}$ & -0.006  & 0.206  & 0.216 &  & -0.006  & 0.145  & 0.146  &  & -0.005  & 0.102  & 0.102  \\
                       &                      & $\theta_{21,0}$ & 0.005   & 0.213  & 0.228 &  & 0.002   & 0.150  & 0.157  &  & 0.002   & 0.105  & 0.106  \\
                       &                      & $\theta_{22,0}$ & -0.009  & 0.214  & 0.237 &  & -0.012  & 0.150  & 0.154  &  & -0.003  & 0.105  & 0.102  \\

                      &                      &       &         &        &       &  &         &        &        &  &         &        &       \\
\multirow{8}{*}{DGP 14} & \multirow{4}{*}{MDD} & $\theta_{11,0}$ & 0.000  & 0.124  & 0.134 &  & 0.001  & 0.092  & 0.093  &  & 0.000  & 0.067  & 0.068  \\
                       &                      & $\theta_{12,0}$ & -0.001  & 0.124  & 0.129 &  & -0.001  & 0.092  & 0.094  &  & -0.000  & 0.067  & 0.067  \\
                       &                      & $\theta_{21,0}$ & 0.001   & 0.126  & 0.133 &  & 0.001   & 0.094  & 0.103  &  & 0.005   & 0.068  & 0.068  \\
                       &                      & $\theta_{22,0}$ & 0.002  & 0.126  & 0.136 &  & -0.002  & 0.094  & 0.100  &  & 0.001  & 0.068  & 0.068  \\
                       & \multirow{4}{*}{DL}  & $\theta_{11,0}$ & 0.001  & 0.185  & 0.199 &  & 0.000  & 0.137  & 0.144  &  & 0.001  & 0.099  & 0.103  \\
                       &                      & $\theta_{12,0}$ & -0.002  & 0.186  & 0.195 &  & -0.003  & 0.136  & 0.138  &  & -0.003  & 0.099  & 0.099  \\
                       &                      & $\theta_{21,0}$ & -0.000   & 0.191  & 0.206 &  & -0.001   & 0.142  & 0.151  &  & 0.001   & 0.102  & 0.103  \\
                       &                      & $\theta_{22,0}$ & -0.006  & 0.192  & 0.212 &  & -0.008  & 0.141  & 0.148  &  & -0.002  & 0.102  & 0.101  \\

                      &                      &       &         &        &       &  &         &        &        &  &         &        &       \\
\multirow{8}{*}{DGP 15} & \multirow{4}{*}{MDD} & $\theta_{11,0}$ & 0.001  & 0.141  & 0.154 &  & 0.001  & 0.099  & 0.106  &  & 0.001  & 0.070  & 0.076  \\
                       &                      & $\theta_{12,0}$ & -0.002  & 0.140  & 0.145 &  & -0.001  & 0.098  & 0.103  &  & -0.001  & 0.069  & 0.071  \\
                       &                      & $\theta_{21,0}$ & 0.003   & 0.143  & 0.154 &  & 0.003   & 0.102  & 0.111  &  & 0.006   & 0.072  & 0.075  \\
                       &                      & $\theta_{22,0}$ & 0.001  & 0.142  & 0.155 &  & -0.003  & 0.100  & 0.107  &  & -0.001  & 0.071  & 0.072  \\
                       & \multirow{4}{*}{DL}  & $\theta_{11,0}$ & 0.003  & 0.209  & 0.247 &  & -0.000  & 0.148  & 0.172  &  & 0.002  & 0.104  & 0.121  \\
                       &                      & $\theta_{12,0}$ & -0.006  & 0.207  & 0.230 &  & -0.006  & 0.145  & 0.155  &  & -0.005  & 0.102  & 0.108  \\
                       &                      & $\theta_{21,0}$ & 0.006   & 0.217  & 0.257 &  & 0.002   & 0.153  & 0.177  &  & 0.002   & 0.107  & 0.120  \\
                       &                      & $\theta_{22,0}$ & -0.011  & 0.215  & 0.250 &  & -0.012  & 0.151  & 0.163  &  & -0.004  & 0.106  & 0.108  \\

                       &                      &           &         &        &       &  &         &        &        &  &         &        &        \\
\multirow{8}{*}{DGP 16} & \multirow{4}{*}{MDD}                  & $\theta_{11,0}$ & -0.017  & 0.090  & 0.094 &  & -0.007  & 0.064  & 0.064  &  & -0.003  & 0.045  & 0.043  \\
                       &                      & $\theta_{12,0}$ & -0.007  & 0.090  & 0.092 &  & -0.002  & 0.064  & 0.065  &  & -0.003  & 0.045  & 0.045  \\
                       &                      & $\theta_{21,0}$ & 0.009   & 0.118  & 0.120 &  & 0.008   & 0.083  & 0.084  &  & 0.005   & 0.059  & 0.058  \\
                       &                      & $\theta_{22,0}$ & -0.019  & 0.116  & 0.118 &  & -0.011  & 0.083  & 0.083  &  & -0.006  & 0.059  & 0.060  \\
                       & \multirow{4}{*}{DL}                   & $\theta_{11,0}$ & -0.006  & 0.142  & 0.143 &  & -0.003  & 0.099  & 0.101  &  & -0.001  & 0.070  & 0.068  \\
                       &                      & $\theta_{12,0}$ & -0.029  & 0.143  & 0.151 &  & -0.013  & 0.099  & 0.101  &  & -0.008  & 0.069  & 0.071  \\
                       &                      & $\theta_{21,0}$ & 0.001   & 0.189  & 0.197 &  & 0.003   & 0.131  & 0.133  &  & 0.001   & 0.091  & 0.093  \\
                       &                      & $\theta_{22,0}$ & -0.043  & 0.186  & 0.200 &  & -0.025  & 0.130  & 0.132  &  & -0.012  & 0.091  & 0.095  \\
                 \bottomrule
\end{tabular}
\end{table}

\section{Empirical Analysis}\label{sec_5}

This section provides two real examples to demonstrate the importance of our proposed estimators.
As before, the estimates from our two-step estimation method and the estimation method in  \citet{DL:2004} are
referred to as ``MDD'' and ``DL'' estimates, respectively.

\subsection{A Univariate Example}
In this subsection, we re-study the data set in \cite{LLZ:2016}.
This data set contains weekly closing prices of Hang Seng Index (HSI) from January 2000 to
December 2007 with 418 observations in total. Define log-return (in percentage) $Y_t=100(\log P_t-\log P_{t-1})$,
where $P_t$ is the closing price of HSI at time $t$. \cite{LLZ:2016} find that
both conditional mean and variance of $y_t$ have the threshold effect, and they fit the conditional mean of
$Y_t$ by an order 2 threshold autoregressive (TAR(2)) model with $Y_{t-1}$ being the threshold variable and $0$ being the threshold.
Following their findings,  we apply our two-step estimation method (with the conditioning variable $X_t=(Y_{t-1},...,Y_{t-4})^{\top}$) to obtain the fitted TAR(2) model below:
\begin{equation} \label{eamdd}
Y_t=\left\{\begin{array}{lc}
-0.608-0.391 Y_{t-1}+0.225 Y_{t-2}+\varepsilon_t , & \text { if } Y_{t-1} \leq 0, \\
(0.520)~(0.218) ~~~~~~~~~ (0.134) &  \\
0.608+0.012 Y_{t-1}-0.094 Y_{t-2}+\varepsilon_t , & \text { if } Y_{t-1}>0, \\
(0.521) ~(0.208) ~~~~~(0.101) &
\end{array}\right.
\end{equation}
where the standard errors of MDD estimates are given in parentheses.
From model (\ref{eamdd}), we find that only the parameter for $Y_{t-2}$ in the regime with respect to $Y_{t-1}\leq 0$ is barely significantly different from zero at the 10\% level. Note that the fitting result in \cite{LLZ:2016} gives a much stronger evidence for this significant parameter. However, their result highly relies on a specific conditionally heteroskedastic model of $Y_t$, whereas our result is robust allowing the conditional heteroskedasticity of $Y_t$ to have an unspecified form.

As a comparison, we also use the DL estimation method in \citet{DL:2004} to get the following fitted TAR(2) model:
\begin{equation} \label{eaDL}
Y_t=\left\{\begin{array}{lc}
-0.974-0.346Y_{t-1}-0.079 Y_{t-2}+\varepsilon_t , & \text { if } Y_{t-1} \leq 0, \\
(0.597)~(0.272) ~~~~~~~(0.252) &  \\
-0.043+0.022 Y_{t-1}-0.018 Y_{t-2}+\varepsilon_t, & \text { if } Y_{t-1}>0, \\
(0.559) \quad(0.349) ~~~~~(0.231) &
\end{array}\right.
\end{equation}
where the same conditioning variable $X_t$ is adopted as for our two-step estimation method, and the standard errors of DL estimates are given in parentheses. Compared with the results in model (\ref{eamdd}),  the DL estimation is less efficient than the two-step estimation (as evidenced by the larger standard errors associated with the DL estimates in (\ref{eaDL})), and it can not detect any significant parameter.

\subsection{A Multivariate Example}

In this subsection, we re-visit a benchmark data set in \cite{Tsay:2005}, which consists of daily log returns of the SP500 index, the stock price of Cisco Systems, and the stock price of Intel Corporation from January 2, 1991 to December 31, 1999 with 2275 observations in total.
We denote this 3-dimensional multivariate
time series by $Y_t = (Y_{1t}, Y_{2t}, Y_{3t})^{\top}$. \cite{Tsay:2005} fits $Y_t$ via the following order 3 vector AR model:
\begin{equation} \label{varcccgarch}
Y_t=A_0+A_1 Y_{t-1}+A_2 Y_{t-2}+A_3 Y_{t-3}+\varepsilon_t,
\end{equation}
where the above model is estimated by the least squares (LS) estimation method. However, since $\varepsilon_t$ has the conditional heteroskedasticity effect as shown in \cite{Tsay:2005}, the standard errors of LS estimates are not reliable under the i.i.d. assumption of $\varepsilon_t$.
To overcome this difficulty, we employ our two-step estimation method (with the conditioning variable $X_t=(Y_{t-1}^{\top},Y_{t-2}^{\top},Y_{t-3}^{\top})^{\top}$) to estimate model (\ref{varcccgarch}), and present the corresponding MDD estimates in Table \ref{eatsay}. For comparison, the DL estimates (with the same conditioning variable $X_t$ as for the two-step estimation method) are also reported in this table.
From Table \ref{eatsay}, we find that (i) the MDD estimates have much smaller values of standard errors than the DL estimates, lending a support that the MDD estimates are much more accurate than the DL estimates; and (ii) the MDD estimates are able to detect much more significant parameters in model (\ref{varcccgarch}) than the DL estimates. Hence, the above findings illustrate the importance of non-integrable weighting function used by the two-step estimation method.

Overall, although both two-step and DL estimation methods allow for an unspecified form of conditional heteroskedasticity, our two real examples show that the former method could provide more accurate estimates than the latter one, and this advantage is more substantial when the data are multivariate.

\begin{table}[!ht]
\caption{The values of MDD and DL estimates and their standard errors.}
\label{eatsay}
\setlength{\tabcolsep}{1.6mm}
\begin{threeparttable}
\begin{tabular}{ccccccc}
\toprule

Parameter           & \multicolumn{3}{c}{MDD$_{(\text{standard error})}$}        & \multicolumn{3}{c}{DL$_{(\text{standard error})}$}        \\

 \cmidrule(lr){1-7}

$A_0$                  & \pmb{0.001}$_{(0.000)}$  & \pmb{0.003}$_{(0.001)}$  & \pmb{0.002}$_{(0.001)}$  & 0.000$_{(0.001)}$  & 0.002$_{(0.001)}$  & 0.001$_{(0.001)}$  \\

\cmidrule(lr){1-7}

\multirow{3}{*}{$A_1$} & 0.012$_{(0.033)}$  & \pmb{0.017}$_{(0.008)}$  & -0.010$_{(0.009)}$ & -0.050$_{(0.120)}$ & 0.022$_{(0.017)}$  & 0.003$_{(0.022)}$  \\
                    & -0.031$_{(0.097)}$ & 0.025$_{(0.030)}$  & 0.041$_{(0.029)}$  & -0.457$_{(0.264)}$ & 0.065$_{(0.056)}$  & \pmb{0.161}$_{(0.062)}$  \\
                    & \pmb{-0.193}$_{(0.086)}$ & -0.002$_{(0.023)}$ & \pmb{0.061}$_{(0.026)}$  & -0.276$_{(0.262)}$ & -0.047$_{(0.055)}$ & \pmb{0.123}$_{(0.057)}$  \\

\cmidrule(lr){1-7}

\multirow{3}{*}{$A_2$} & 0.015$_{(0.034)}$  & -0.006$_{(0.008)}$ & 0.001$_{(0.009)}$  & 0.001$_{(0.119)}$  & 0.019$_{(0.017)}$  & -0.020$_{(0.025)}$ \\
                    & \pmb{0.306}$_{(0.095)}$  & \pmb{-0.107}$_{(0.030)}$ & -0.033$_{(0.029)}$ & 0.088$_{(0.276)}$  & -0.084$_{(0.063)}$ & -0.067$_{(0.073)}$ \\
                    & -0.012$_{(0.082)}$ & -0.005$_{(0.021)}$ & -0.005$_{(0.027)}$ & -0.158$_{(0.222)}$ & \pmb{0.128}$_{(0.054)}$  & -0.120$_{(0.068)}$ \\

\cmidrule(lr){1-7}

\multirow{3}{*}{$A_3$} & \pmb{-0.092}$_{(0.030)}$ & 0.005$_{(0.007)}$  & 0.009$_{(0.009)}$  & -0.153$_{(0.085)}$ & -0.021$_{(0.021)}$ & 0.045$_{(0.024)}$  \\
                    & -0.079$_{(0.092)}$ & -0.038$_{(0.030)}$ & 0.026$_{(0.029)}$  & -0.278$_{(0.226)}$ & -0.121$_{(0.062)}$ & 0.070$_{(0.066)}$  \\
                    & -0.037$_{(0.081)}$ & -0.004$_{(0.022)}$ & -0.021$_{(0.027)}$ & 0.088$_{(0.208)}$  & -0.090$_{(0.053)}$ & -0.042$_{(0.059)}$ \\ \bottomrule
\end{tabular}
 \begin{tablenotes}
   \item {\textit{Note}: The estimates in boldface are significantly different from zero at the 5\% level.}
 \end{tablenotes}
\end{threeparttable}
\end{table}

\section{Concluding Remarks}\label{sec_6}

In this paper, we propose a new MDD-based estimation method for conditional moment models.
This MDD-based estimation method roots in the idea of a continuum of unconditional moment restrictions, with a non-integrable weighting function to ensemble the information from different estimating moment functions. Under conditions that allow for time series data with conditional heteroskedasticity of unknown form, the proposed estimators by our MDD-based method are shown to be asymptotically normal with analytic asymptotic variances. Hence, all of our prosed estimators are easy for implementing statistical inference and have a large application scope for studying multiple linear/nonlinear time series models. Their importance is intensively demonstrated through simulations and two real examples.
As a future work, it is interesting to extend our MDD-based estimation idea to the non-smooth function $h$ as in \cite{CP:2009,CP:2012} and \cite{GS:2012}, and this may call for non-trivial technical treatments.

\section*{Acknowledgments}
The codes and data used for this paper are accessible at ``\url{https://github.com/hkusky/MDDEstimation.git}''.

\renewcommand{\thesection}{A}
\setcounter{equation}{0}
\setcounter{section}{0}
\section*{Appendix: Proofs of Theorems \ref{thm1}--\ref{thm2} and \ref{thm3}}

To facilitate the proofs of all theorems, we let $C$ be a generic constant whose value changes from
place to place and introduce the following additional notation:
\begin{equation}\label{eqn_A_1}
\begin{split}
&h_t(\theta)= h(Z_t,\theta),\,\,\,\dot{h}_t(\ct)= \frac{\partial h(Z_t,\ct)}{\partial\theta},\,\,\,\dot{\mathcal{G}}_n(s,\ct)=\frac{\partial \mathcal{G}_n(s,\ct)}{\partial \ct},\\
&\mathcal{G}(s,\ct)=E\big\{[h_t(\ct)-E(h_t(\ct))]\exp(\mathrm{i} \langle s, X_t\rangle)\big\},\\
&\dot{\mathcal{G}}(s,\ct)=E\big\{[\dot{h}_t(\ct)-E(\dot{h}_t(\ct))]\exp(\mathrm{i} \langle s, X_t\rangle)\big\}.
\end{split}
\end{equation}
Below, we give a technical lemma for proving Theorem \ref{thm1}.

\begin{lem}\label{lem_A1}
Denote
$$\mathcal{G}_n(s,\theta)^{\star}\mathcal{G}_n(s,\theta)
=C_{1n}(s,\theta)+\textrm{i}C_{2n}(s,\theta)\,\,\,\mbox{ and }\,\,\,
\mathcal{G}(s,\theta)^{\star}\mathcal{G}(s,\theta)
=C_{3}(s,\theta)+\textrm{i}C_{4}(s,\theta)$$
for any $\theta\in\Theta$, where $\mathcal{G}_n(s,\ct)$ and $\mathcal{G}(s,\theta)$ are defined in (\ref{g_n_s_theta}) and (\ref{eqn_A_1}), respectively,
and $C_{1n}(s,\theta)$, $C_{2n}(s,\theta)$, $C_{3}(s,\theta)$, and $C_{4}(s,\theta)$ are four scalars.
Suppose Assumptions \ref{stationarity}--\ref{c2} hold. Then, uniformly in $\theta\in\Theta$,

(i) $E\|C_{1n}(s,\theta)\|\leq C(1\land\|s\|^2)$ and $E\|C_{2n}(s,\theta)\|\leq C(1\land\|s\|^2)$ for all $s\in\mathbb{R}^{q}$;

(ii) $\|C_{3}(s,\theta)\|\leq C(1\land\|s\|^2)$ and $\|C_{4}(s,\theta)\|\leq C(1\land\|s\|^2)$ for all $s\in\mathbb{R}^{q}$.
\end{lem}

\noindent \textbf{Proof of Lemma \ref{lem_A1}.} The results (i) and (ii) hold by using the similar arguments as for Lemma 6 in the supplementary materials of \cite{WZS:2022}. \qed

\vspace{2mm}

\noindent \textbf{Proof of Theorem \ref{thm1}.}
We use a similar truncation technique as in \cite{Davis:2018} to facilitate the proof.
Define a compact set
\begin{flalign}\label{compact_set}
\Pi_\rho=\{s\in\mathbb{R}^q:\rho\leq \|s\|\leq 1/\rho\} \mbox{ for some }\rho>0.
\end{flalign}
By Assumptions \ref{stationarity}--\ref{c2} and the uniformly ergodic theorem in \cite{J:1969}, we have
\begin{flalign}\label{eqn_A_3}
\mathcal{G}_n(s,\theta) \Longrightarrow \mathcal{G}(s,\theta) \mbox{ on }\Pi_{\rho}\times \Theta \mbox{ as }n\to\infty,
\end{flalign}
where ``$\Longrightarrow$'' stands for the weak convergence, and  $\mathcal{G}_n(s,\theta)$ and $ \mathcal{G}(s,\ct)$ are defined in (\ref{g_n_s_theta}) and (\ref{eqn_A_1}), respectively.  We note that $ \mathcal{G}(s,\ct)$ is deterministic. Hence, by the continuous mapping theorem and (\ref{eqn_A_3}), we can obtain that $\int_{\Pi_\rho}\Big|\mathcal{G}_n(s,\theta)\Big|_{l}^{2}w_*(s)ds \Longrightarrow \int_{\Pi_\rho} \Big| \mathcal{G}(s,\theta)\Big|_{l}^2 w_*(s)ds$
on $\Theta$, indicating
\begin{flalign}\label{eqn_A_4}
\sup_{\theta\in\Theta}\Big\|\int_{\Pi_\rho}\Big[\Big|\mathcal{G}_n(s,\theta)\Big|_{l}^{2}-\Big| \mathcal{G}(s,\theta)\Big|_{l}^2\Big]w_*(s)ds\Big\|
\overset{p}{\longrightarrow}0 \mbox{ as }n\to\infty.
\end{flalign}

Next, by Lemma \ref{lem_A1}(i), the Markov inequality, and the fact $\int_{\mathbb{R}^q} (1\land\|s\|^2) w_*(s)ds<\infty$, it is straightforward to see that uniformly in $\theta\in\Theta$,
\begin{flalign}\label{eqn_A_5}
\mathop{\lim}\limits_{\rho \to 0} \mathop{\limsup}\limits_{n\rightarrow \infty} \int_{\Pi_\rho^c}\Big|\mathcal{G}_n(s,\theta)\Big|_{l}^{2}w_*(s)ds=0,
\end{flalign}
in probability. Similarly, by Lemma \ref{lem_A1}(ii) we have
\begin{flalign}\label{eqn_A_6}
\mathop{\lim}\limits_{\rho \to 0} \mathop{\limsup}\limits_{n\rightarrow \infty} \int_{\Pi_\rho^c} \Big|\mathcal{G}(s,\theta)\Big|_{l}^2w_*(s)ds=0.
\end{flalign}
Therefore, by (\ref{eqn_A_4})--(\ref{eqn_A_6}), it follows that
\begin{flalign*}
\sup_{\theta\in\Theta}\Big\|\int_{\mathbb{R}^q}\Big[\Big|\mathcal{G}_n(s,\theta)\Big|_{l}^{2}-\Big|\mathcal{G}(s,\theta)\Big|_{l}^2\Big]w_*(s)ds\Big\|
\overset{p}{\longrightarrow}0 \mbox{ as }n\to\infty,
\end{flalign*}
which is equivalent to
\begin{equation}\label{eqn_A_7}
\mathop{\sup}\limits_{\ct \in \CT}\big\|\operatorname{MDD}_n(\ct)-\operatorname{MDD}(\ct)\big\|\overset{p}{\longrightarrow} 0 \mbox{ as }n\to\infty,
\end{equation}
in view of (\ref{eqn1_7}) and (\ref{mdd_sample_integral}).

Furthermore, by Theorem 1 of \cite{SZ:2014}, we know that $\operatorname{MDD}(\ct)\geq0$ and the equality holds if and only if $E\big[h(Z_t,\ct)-E(h(Z_t,\ct))|X_t\big]=0$ (a.s.). Hence, $\operatorname{MDD}(\ct)$ has a unique minimum point $\ct_0$ by Assumption \ref{hatii}.
Now, the conclusion holds by (\ref{eqn_A_7}), Assumption \ref{interpoint}, and the standard arguments.
\qed

\vspace{2mm}

To prove Theorems \ref{thm2} and \ref{thm3}, we need three additional lemmas below. The first lemma is analogous to Lemma \ref{lem_A1},
the second lemma is to show the uniform boundedness of $\sqrt{n}{\mathcal{G}}_n(s,\ct_0)$ in probability,
and the third lemma is key to obtain the analytic forms of asymptotic variance in both theorems.

\begin{lem}\label{lem_A2}
Let $\widehat{\theta}_{n}^{\,'}\in\Theta$ and $\widehat{\theta}_{n}^{\,''}\in\Theta$ be any two estimators of $\theta_0$. Denote
$$\dot{\mathcal{G}}_n(s,\widehat{\theta}_{n}^{\,'})^{\star}\dot{\mathcal{G}}_n(s,\widehat{\theta}_{n}^{\,''})
=C_{5n}(s)+\textrm{i}C_{6n}(s)\,\,\,\mbox{ and }\,\,\,
\dot{\mathcal{G}}_n(s,\widehat{\theta}_{n}^{\,'})^{\star}[\sqrt{n}\mathcal{G}_n(s,\theta_{0})]=C_{7n}(s)+\textrm{i}C_{8n}(s),$$
where $\mathcal{G}_n(s,\theta)$ and $\dot{\mathcal{G}}_n(s,\ct)$ are defined in (\ref{g_n_s_theta}) and (\ref{eqn_A_1}), respectively,
$C_{5n}(s)$ and $C_{6n}(s)$ are two $d\times d$ matrices, and $C_{7n}(s)$ and $C_{8n}(s)$ are two $d$-dimensional vectors.
Suppose Assumptions \ref{stationarity}--\ref{c2} and \ref{whitenoise} hold. Then,

(i) $E\|c_{5n,ij}(s)\|\leq C(1\land\|s\|^2)$ and $E\|c_{6n,ij}(s)\|\leq C(1\land\|s\|^2)$ for all $s\in\mathbb{R}^{q}$,

\noindent where $c_{5n,ij}(s)$ and $c_{6n,ij}(s)$ are the $(i,j)$th entry of $C_{5n}(s)$ and $C_{6n}(s)$, respectively;

(ii) $E\|c_{7n,i}(s)\|\leq C(1\land\|s\|^2)$ and $E\|c_{8n,i}(s)\|\leq C(1\land\|s\|^2)$ for all $s\in\mathbb{R}^{q}$,

\noindent where $c_{7n,i}(s)$ and $c_{8n,i}(s)$ are the $i$th entry of $C_{7n}(s)$ and $C_{8n}(s)$, respectively.
\end{lem}

\noindent \textbf{Proof of Lemma \ref{lem_A2}.} The results (i) and (ii) hold by using the similar arguments as for Lemma 6 in the supplementary materials of \cite{WZS:2022}. \qed

\begin{lem}\label{lem_A3}
Suppose Assumptions \ref{stationarity}--\ref{c2} and \ref{whitenoise} hold. Then,
$$\sup_{s\in\Pi}\Big\|\sqrt{n}{\mathcal{G}}_n(s,\ct_0)\Big\|=O_{p}(1),$$
for any compact set $\Pi\subset \mathbb{R}^{q}$.
\end{lem}

\noindent \textbf{Proof of Lemma \ref{lem_A3}.} By using the similar arguments as for Lemma 4 in the supplementary materials of \cite{WZS:2022},
we can show that on $\Pi$, $\sqrt{n}{\mathcal{G}}_n(s,\ct_0)$ converges weakly to a Gaussian process $\chi(s)$ as $n\to\infty$. Then, the conclusion follows directly. \qed

\begin{lem}\label{lem_A4}
Suppose Assumptions \ref{stationarity}--\ref{c2} and \ref{whitenoise} hold. Then,

(i) $\int_{\mathbb{R}^q}\dot{\mathcal{G}}(s,\ct_0)^{\star}\dot{\mathcal{G}}(s,\ct_0)w_*(s)ds=-\Omega$;

(ii) $\int_{\mathbb{R}^q}\dot{\mathcal{G}}(s,\ct_0) ^{\star} [\sqrt{n}\mathcal{G}_n(s,\theta_0)]w_*(s)ds=-n^{-1/2}
\sum_{t=1}^{n}[u(X_t)-E(u(X_t))]^{\top}h_t(\theta_0)+o_{p}(1)$,

\noindent where $\mathcal{G}_n(s,\theta)$ and $\dot{\mathcal{G}}_n(s,\ct)$ are defined in (\ref{g_n_s_theta}) and (\ref{eqn_A_1}), respectively,
and $u(x)$ is defined in (\ref{def_u}).
\end{lem}


\noindent \textbf{Proof of Lemma \ref{lem_A4}.} (i) Re-write
$$\dot{\mathcal{G}}(s,\ct_0)=E\big\{[\dot{h}_{t'}(\ct_0)-E(\dot{h}_{t'}(\ct_0))]\exp(\mathrm{i} \langle s, X_{t'}\rangle)\big\},$$
where  $(\dot{h}_{t'}(\ct_0),X_{t'})$ is an i.i.d. copy of
$(\dot{h}_{t}(\ct_0),X_{t})$. Due to the independence of $(\dot{h}_{t'}(\ct_0),X_{t'})$ and $(\dot{h}_{t}(\ct_0),X_{t})$,
it follows that
\begin{flalign}\label{eqn_A_8}
\int_{\mathbb{R}^q}\dot{\mathcal{G}}(s,\ct_0)^{\star}\dot{\mathcal{G}}(s,\ct_0)w_*(s)ds &=\int_{\mathbb{R}^q}E\big\{[\dot{h}_{t'}(\ct_0)-E(\dot{h}_{t'}(\ct_0))]^{\top}\exp(-\mathrm{i} \langle s, X_{t'}\rangle)\big\} \nonumber\\
&\quad \times E\big\{[\dot{h}_{t}(\ct_0)-E(\dot{h}_{t}(\ct_0))]\exp(\mathrm{i} \langle s, X_{t}\rangle)\big\}w_*(s)ds \nonumber\\
&=\int_{\mathbb{R}^q}E\big\{[\dot{h}_{t'}(\ct_0)-E(\dot{h}_{t'}(\ct_0))]^{\top}[\dot{h}_{t}(\ct_0)-E(\dot{h}_{t}(\ct_0))] \nonumber\\
&\quad \times \exp(\mathrm{i} \langle s, X_{t}-X_{t'}\rangle)\big\}w_*(s)ds \nonumber\\
&=\int_{\mathbb{R}^q}E\big\{[\dot{h}_{t'}(\ct_0)-E(\dot{h}_{t'}(\ct_0))]^{\top}[\dot{h}_{t}(\ct_0)-E(\dot{h}_{t}(\ct_0))] \nonumber\\
&\quad \times [\cos(\langle s, X_{t}-X_{t'}\rangle)-1]\big\}w_*(s)ds \nonumber\\
&=-E\big\{[\dot{h}_{t'}(\ct_0)-E(\dot{h}_{t'}(\ct_0))]^{\top}[\dot{h}_{t}(\ct_0)-E(\dot{h}_{t}(\ct_0))] \nonumber\\
&\quad \times\|X_{t}-X_{t'}\|\big\},
\end{flalign}
where the next-to-last result holds since
 $E\big\{[\dot{h}_{t'}(\ct_0)-E(\dot{h}_{t'}(\ct_0))]^{\top}[\dot{h}_{t}(\ct_0)-E(\dot{h}_{t}(\ct_0))]\big\}=0$ and $\sin(\cdot)$ is an odd function, and the last result holds by the identity (\ref{identity}).

Let $E_{t}$ denote the expectation over $(\dot{h}_{t}(\ct_0), X_t)$. Then, by (\ref{eqn_A_8}), the properties of double expectation, and the
independence of $(\dot{h}_{t'}(\ct_0),X_{t'})$ and $(\dot{h}_{t}(\ct_0),X_{t})$, we have
\begin{flalign*}
&\int_{\mathbb{R}^q}\dot{\mathcal{G}}(s,\ct_0)^{\star}\dot{\mathcal{G}}(s,\ct_0)w_*(s)ds\\
&=-E\big\{[\dot{h}_{t'}(\ct_0)-E(\dot{h}_{t'}(\ct_0))]^{\top}E\big\{[\dot{h}_{t}(\ct_0)-E(\dot{h}_{t}(\ct_0))]\|X_{t}-X_{t'}\|\big| \dot{h}_{t'}(\ct_0), X_{t'}\big\}\big\}\\
&=-E\big\{[\dot{h}_{t'}(\ct_0)-E(\dot{h}_{t'}(\ct_0))]^{\top}E_{t}\big\{[\dot{h}_{t}(\ct_0)-E(\dot{h}_{t}(\ct_0))]\|X_{t}-X_{t'}\|\big| \dot{h}_{t'}(\ct_0), X_{t'}\big\}\big\}\\
&=-E\big\{[\dot{h}_{t'}(\ct_0)-E(\dot{h}_{t'}(\ct_0))]^{\top}E_{t}\big\{[\dot{h}_{t}(\ct_0)-E(\dot{h}_{t}(\ct_0))]\|X_{t}-X_{t'}\|\big\}\big\}\\
&=-E\big\{[\dot{h}_{t'}(\ct_0)-E(\dot{h}_{t'}(\ct_0))]^{\top}u(X_{t'})\big\}\\
&=-\Omega.
\end{flalign*}
Hence, the result (i) holds.

(ii) Note that $u(X_t)=E_{t'}\big\{[\dot{h}_{t'}(\ct_0)-E(\dot{h}_{t'}(\ct_0))]\|X_{t}-X_{t'}\|\big\}$. By using the similar arguments as for (\ref{eqn_A_8}), we have
\begin{flalign}\label{eqn_A_9}
&\frac{1}{\sqrt{n}}\sum_{t=1}^{n}u(X_t)^{\top}p_t(\theta_0) \nonumber\\
&=\frac{1}{\sqrt{n}}\sum_{t=1}^{n}
E_{t'}\big\{[\dot{h}_{t'}(\ct_0)-E(\dot{h}_{t'}(\ct_0))]^{\top}\|X_{t}-X_{t'}\|\big\}p_t(\theta_0) \nonumber\\
&=\int_{\mathbb{R}^q} \frac{1}{\sqrt{n}}\sum_{t=1}^{n}
E_{t'}\big\{[\dot{h}_{t'}(\ct_0)-E(\dot{h}_{t'}(\ct_0))]^{\top}[1-\cos(\langle s, X_{t}-X_{t'}\rangle)]\big\}p_t(\theta_0)w_{*}(s)ds \nonumber\\
&=-\int_{\mathbb{R}^q} \frac{1}{\sqrt{n}}\sum_{t=1}^{n}
E_{t'}\big\{[\dot{h}_{t'}(\ct_0)-E(\dot{h}_{t'}(\ct_0))]^{\top}\cos(\langle s, X_{t}-X_{t'}\rangle)\big\}p_t(\theta_0)w_{*}(s)ds \nonumber\\
&=-\int_{\mathbb{R}^q} \frac{1}{\sqrt{n}}\sum_{t=1}^{n}
E_{t'}\big\{[\dot{h}_{t'}(\ct_0)-E(\dot{h}_{t'}(\ct_0))]^{\top}\exp(\textrm{i}\langle s, X_{t}-X_{t'}\rangle)\big\}p_t(\theta_0)w_{*}(s)ds,
\end{flalign}
where $p_t(\theta_0)=h_t(\theta_0)-\overline{h}(\theta_0)$ with $\overline{h}(\theta_0)=n^{-1}\sum_{t=1}^{n}h_t(\theta_0)$.

Next, since $(\dot{h}_{t'}(\ct_0),X_{t'})$ and $X_{t}$ are independent, we can obtain
\begin{flalign}\label{eqn_A_10}
&\frac{1}{\sqrt{n}}\sum_{t=1}^{n}
E_{t'}\big\{[\dot{h}_{t'}(\ct_0)-E(\dot{h}_{t'}(\ct_0))]^{\top}\exp(\textrm{i}\langle s, X_{t}-X_{t'}\rangle)\big\}p_t(\theta_0) \nonumber\\
&=\frac{1}{\sqrt{n}}\sum_{t=1}^{n}
E_{t'}\big\{[\dot{h}_{t'}(\ct_0)-E(\dot{h}_{t'}(\ct_0))]^{\top}\exp(\textrm{i}\langle s, -X_{t'}\rangle)\big\}\exp(\textrm{i}\langle s, X_{t}\rangle)p_t(\theta_0) \nonumber\\
&=E_{t'}\big\{[\dot{h}_{t'}(\ct_0)-E(\dot{h}_{t'}(\ct_0))]^{\top}\exp(\textrm{i}\langle s, -X_{t'}\rangle)\big\}\times\Big\{\frac{1}{\sqrt{n}}\sum_{t=1}^{n}p_t(\theta_0)
\exp(\textrm{i}\langle s, X_{t}\rangle)\Big\} \nonumber\\
&=\dot{\mathcal{G}}(s,\ct_0)^{\star}\Big\{\frac{1}{\sqrt{n}}\sum_{t=1}^{n}
p_t(\theta_0)\exp(\textrm{i}\langle s, X_{t}\rangle)\Big\}.
\end{flalign}
Thus, by (\ref{eqn_A_9})--(\ref{eqn_A_10}) and (\ref{g_n_s_theta}), we have
\begin{flalign}\label{eqn_A_11}
\frac{1}{\sqrt{n}}\sum_{t=1}^{n}u(X_t)^{\top}p_t(\theta_0)&=-\int_{\mathbb{R}^q} \dot{\mathcal{G}}(s,\ct_0)^{\star}\Big\{\frac{1}{\sqrt{n}}\sum_{t=1}^{n}
p_t(\theta_0)\exp(\textrm{i}\langle s, X_{t}\rangle)\Big\}w_{*}(s)ds \nonumber\\
&=-\int_{\mathbb{R}^q} \dot{\mathcal{G}}(s,\ct_0)^{\star}[\sqrt{n}\mathcal{G}_n(s,\theta_0)]w_{*}(s)ds.
\end{flalign}

Moreover, it is straightforward to see
\begin{flalign}\label{eqn_A_12}
\frac{1}{\sqrt{n}}\sum_{t=1}^{n}u(X_t)^{\top}p_t(\theta_0)&=\frac{1}{\sqrt{n}}\sum_{t=1}^{n}u(X_t)^{\top}h_t(\theta_0)
-\Big(\frac{1}{n}\sum_{t=1}^{n}u(X_t)^{\top}\Big)\Big(\frac{1}{\sqrt{n}}\sum_{t=1}^{n}h_t(\theta_0)\Big) \nonumber\\
&=\frac{1}{\sqrt{n}}\sum_{t=1}^{n}[u(X_t)-E(u(X_t))]^{\top}h_t(\theta_0)+o_{p}(1),
\end{flalign}
where $n^{-1}\sum_{t=1}^{n}u(X_t)=E(u(X_t))+o_{p}(1)$ by Assumptions \ref{stationarity}--\ref{c2} and the law of large numbers for stationary sequence, and
$n^{-1/2}\sum_{t=1}^{n}h_t(\theta_0)=O_{p}(1)$ by Assumptions \ref{stationarity}--\ref{c2} and \ref{whitenoise},  and the martingale central limit theorem.

Finally, the result (ii) holds by (\ref{eqn_A_11})--(\ref{eqn_A_12}). This completes all of the proof. \qed

\vspace{2mm}

\noindent \textbf{Proof of Theorem \ref{thm2}.} By the definition of $\widehat{\theta}_n$ and \eqref{mdd_sample_integral}, it is straightforward to see
that $\int_{\mathbb{R}^q}\dot{\mathcal{G}}_n(s,\widehat{\ct}_n)^{\star}\mathcal{G}_n(s,\widehat{\ct}_n) w_*(s)ds=0$. Applying the mean value theorem, it follows that
$\int_{\mathbb{R}^q}\big[\dot{\mathcal{G}}_n(s,\widehat{\ct}_n)^{\star}{\mathcal{G}}_n(s,\ct_0)
+\dot{\mathcal{G}}_n(s,\widehat{\ct}_n)^{\star}\dot{\mathcal{G}}_n(s,\widehat{\ct}_n^{\dag})(\widehat{\ct}_n-\ct_0)\big] w_*(s)ds=0$,
where $\widehat{\ct}_{n}^{\dag}$ lies between $\ct_0$ and $\widehat{\ct}_n$. Consequently, we have
\begin{equation}\label{eqn_A_13}
\sqrt{n}(\widehat{\ct}_n-\ct_0)=-A_n^{-1}B_n,
\end{equation}
where $A_n=\int_{\mathbb{R}^q}\dot{\mathcal{G}}_n(s,\widehat{\ct}_n)^{\star} \dot{\mathcal{G}}_n(s,\widehat{\ct}^{\dag})w_*(s)ds$
and $B_n=\int_{\mathbb{R}^q}\dot{\mathcal{G}}_n(s,\widehat{\ct}_n)^{\star}[\sqrt{n}{\mathcal{G}}_n(s,\ct_0)]w_*(s)ds$.

Next, we apply a similar truncation technique as in \cite{Davis:2018} to deal with $A_n$ and $B_n$.
For $A_n$, we re-write it into two parts:
$$A_n=\int_{\Pi_\rho}\dot{\mathcal{G}}_n(s,\widehat{\ct}_n)^{\star} \dot{\mathcal{G}}_n(s,\widehat{\ct}^{\dag})w_*(s)ds
+\int_{\Pi_\rho^{c}}\dot{\mathcal{G}}_n(s,\widehat{\ct}_n)^{\star} \dot{\mathcal{G}}_n(s,\widehat{\ct}^{\dag})w_*(s)ds,$$
where $\Pi_\rho$ is defined in (\ref{compact_set}). On one hand, the first part of $A_n$ satisfies
\begin{flalign}\label{eqn_A_14}
\int_{\Pi_\rho}\dot{\mathcal{G}}_n(s,\widehat{\ct}_n)^{\star} \dot{\mathcal{G}}_n(s,\widehat{\ct}^{\dag})w_*(s)ds
&=\int_{\Pi_\rho}\dot{\mathcal{G}}(s,\widehat{\ct}_n)^{\star} \dot{\mathcal{G}}(s,\widehat{\ct}^{\dag})w_*(s)ds+o_{p}(1) \nonumber\\
&=\int_{\Pi_\rho}\dot{\mathcal{G}}(s,\ct_0)^{\star} \dot{\mathcal{G}}(s,\ct_0)w_*(s)ds+o_{p}(1),
\end{flalign}
where the first equality holds by using the same arguments as for (\ref{eqn_A_3}) and the continuous mapping theorem, and the second equality holds
by Theorem \ref{thm1} and the dominated convergence theorem. On the other hand, the second part of $A_n$ satisfies
\begin{flalign}\label{eqn_A_15}
\mathop{\lim}\limits_{\rho \to 0} \mathop{\limsup}\limits_{n\rightarrow \infty} \int_{\Pi_\rho^{c}}\Big\|\dot{\mathcal{G}}_n(s,\widehat{\ct}_n)^{\star} \dot{\mathcal{G}}_n(s,\widehat{\ct}^{\dag})\Big\|w_*(s)ds=0,
\end{flalign}
in probability, where the result holds by Lemma \ref{lem_A2}(i) and using the same arguments as for (\ref{eqn_A_5}).
Thus, by (\ref{eqn_A_14})--(\ref{eqn_A_15}) we have
\begin{flalign}\label{eqn_A_16}
A_n=\int_{\mathbb{R}^{q}}\dot{\mathcal{G}}(s,\ct_0)^{\star} \dot{\mathcal{G}}(s,\ct_0)w_*(s)ds+o_{p}(1).
\end{flalign}
Similarly, by Lemmas \ref{lem_A2}(ii) and \ref{lem_A3}, we can show
\begin{flalign}\label{eqn_A_17}
B_n=\int_{\mathbb{R}^q}\dot{\mathcal{G}}(s,\ct_0)^{\star}[\sqrt{n}{\mathcal{G}}_n(s,\ct_0)]w_*(s)ds+o_{p}(1).
\end{flalign}

Moreover, by (\ref{eqn_A_13}), (\ref{eqn_A_16})--(\ref{eqn_A_17}), and Lemma \ref{lem_A4}, it follows that
\begin{flalign}\label{eqn_A_18}
\sqrt{n}(\widehat{\ct}_n-\ct_0)=-\Omega^{-1}\Big\{\frac{1}{\sqrt{n}}
\sum_{t=1}^{n}[u(X_t)-E(u(X_t))]^{\top}h_t(\theta_0)\Big\}+o_{p}(1).
\end{flalign}
Finally, the conclusion holds by (\ref{eqn_A_18}), Assumptions \ref{stationarity}--\ref{c2} and \ref{whitenoise}, and the martingale central limit theorem. \qed

\vspace{2mm}

\noindent \textbf{Proof of Theorem \ref{thm3}.} First, as for (\ref{eqn_A_18}), it is not hard to see
\begin{flalign}\label{eqn_A_19}
\sqrt{n}(\widehat{\ct}_{2n}-\ct_{20})=-\Omega_2^{-1}\Big\{\frac{1}{\sqrt{n}}
\sum_{t=1}^{n}[u_2(X_t)-E(u_2(X_t))]^{\top}h_t(\theta_0)\Big\}+o_{p}(1),
\end{flalign}
where $\Omega_2$ and $u_2(x)$ are defined in (\ref{def_u_2}).
Next, by (\ref{special_model})--(\ref{partition}), we have
\begin{align*}
\sqrt{n}(\widehat{\ct}_{1n}-\ct_{10})&=-\frac{1}{\sqrt{n}}\sum_{t=1}^n(m_1(Z_t,\widehat{\ct}_{2n})-m_1(Z_t,\ct_{20})+h_1(Z_t,\ct_0))\\
&=-E\Big(\frac{\partial m_1(Z_t,\theta_{20})}{\partial\theta_2}\Big)[\sqrt{n}(\widehat{\ct}_{2n}-\ct_{20})]-\frac{1}{\sqrt{n}}\sum_{t=1}^n h_1(Z_t,\ct_0)+o_p(1),
\end{align*}
where the last result holds by Taylor's expansion, the uniformly ergodic theorem, and the dominated convergence theorem.
By (\ref{eqn_A_19}), it follows that
\begin{align}\label{eqn_A_20}
\sqrt{n}(\widehat{\ct}_{1n}-\ct_{10})&=\frac{1}{\sqrt{n}}\sum_{t=1}^n
\Big\{E\Big(\frac{\partial m_1(Z_t,\theta_{20})}{\partial\theta_2}\Big) \Omega_2^{-1} [u_2(X_t)-E(u_2(X_t))]^{\top}-\Upsilon\Big\}h_t(\theta_0) \nonumber\\
&\quad +o_p(1),
\end{align}
where we have used the fact that $h_1(Z_t,\ct_0)=\Upsilon h_t(\theta_0)$ with $\Upsilon$ defined in (\ref{def_u_2}). Now, the conclusion holds
by (\ref{eqn_A_19})--(\ref{eqn_A_20}), Assumptions \ref{stationarity}--\ref{c2} and \ref{whitenoise}, and the martingale central limit theorem.
\qed

\end{document}